\documentclass[twocolumn,times]{aastex61}
\turnoffedit
\usepackage{amsmath,amstext}
\usepackage[figure,figure*]{hypcap}
\usepackage{xspace}
\usepackage{graphicx}
\usepackage{color}
\usepackage{tikz}
\usetikzlibrary{shapes.geometric, arrows}
\shorttitle{HST WFC3 Ramp Effect Model}
\shortauthors{Zhou et al.}
\newcommand{\glenn}{}
\begin{document}

\title{A Physical Model-based Correction for Charge Traps in the
  Hubble Space Telescope's Wide Field Camera 3 Near-IR Detector and Applications to
  Transiting Exoplanets and Brown Dwarfs}

\correspondingauthor{Yifan Zhou}
\email{yzhou@as.arizona.edu}

\author[0000-0003-2969-6040]{Yifan Zhou}
\altaffiliation{NASA Earth and Space Science Fellow}
\affil{Department of Astronomy/Steward Observatory, The University of Arizona, 933 N. Cherry Ave., Tucson, AZ, 85721, USA}

\author[0000-0003-3714-5855]{D\'aniel Apai}
\affiliation{Department of Astronomy/Steward Observatory, The
  University of Arizona, 933 N. Cherry Ave., Tucson, AZ, 85721, USA}
\affiliation{Department of Planetary Science/Lunar and Planetary
  Laboratory, The University of Arizona, 1640 E. University Blvd.,
  Tucson, AZ 85718, USA}
\affiliation{Earths in Other Solar Systems
  Team, NASA Nexus for Exoplanet System Science, 933 N. Cherry Ave., Tucson, AZ, 85721, USA}

\author[0000-0003-1487-6452]{Ben W. P. Lew}
\affiliation{Department of Planetary Science/Lunar and Planetary
  Laboratory, The University of Arizona, 1640 E. University Blvd.,
  Tucson, AZ 85718, USA}

\author{Glenn Schneider}
\affiliation{Department of Astronomy/Steward Observatory, The
  University of Arizona, 933 N. Cherry Ave., Tucson, AZ, 85721, USA}

\newcommand{\HST}{\textit{HST}\xspace}
\newcommand{\WFC}{WFC3\xspace}
\newcommand{\ra}{\ensuremath{\to}}
\begin{abstract}
  The Hubble Space Telescope (\HST) Wide Field Camera 3 (\WFC)
  near-IR channel is extensively used in time-resolved
  observations, especially for transiting exoplanet spectroscopy and
  brown dwarf and directly imaged exoplanet rotational phase
  mapping. The \textit{ramp effect} is the dominant source of
  systematics in the \WFC for time-resolved observations, which limits
  its photometric precision. Current mitigation strategies are based
  on empirical fits and require additional orbits ``to help the
  telescope reach a thermal equilibrium.''
  We show that the ramp effect profiles can be explained and corrected
  with high fidelity using charge trapping theories.  We also present
  a model for this process that can be used to predict and to correct
  charge trap systematics.  Our model is based on a very small number
  of parameters that are intrinsic to the detector. We find
  that these parameters are very stable between the different
  datasets, and we provide best-fit values.  Our model is tested with
  more than 120 orbits ($\sim40$ visits) of \WFC observations,
  and is proved to be able to provide near photon noise limited
  corrections for observations made with both staring and scanning
  modes of transiting exoplanets as well as for starting-mode
  observations of brown dwarfs. After our model correction, the light
  curve of the first orbit in each visit has the same photometric
  precision as subsequent orbits, so data from the first orbit need no
  longer be discarded.  Near IR arrays with the same physical
  characteristics (e.g., \textit{JWST/NIRCam}) may also
  benefit from the extension of this model if similar
    systematic profiles are observed.
\end{abstract}


\section{Introduction}

\begin{figure*}[t]
  \centering
  \plottwo{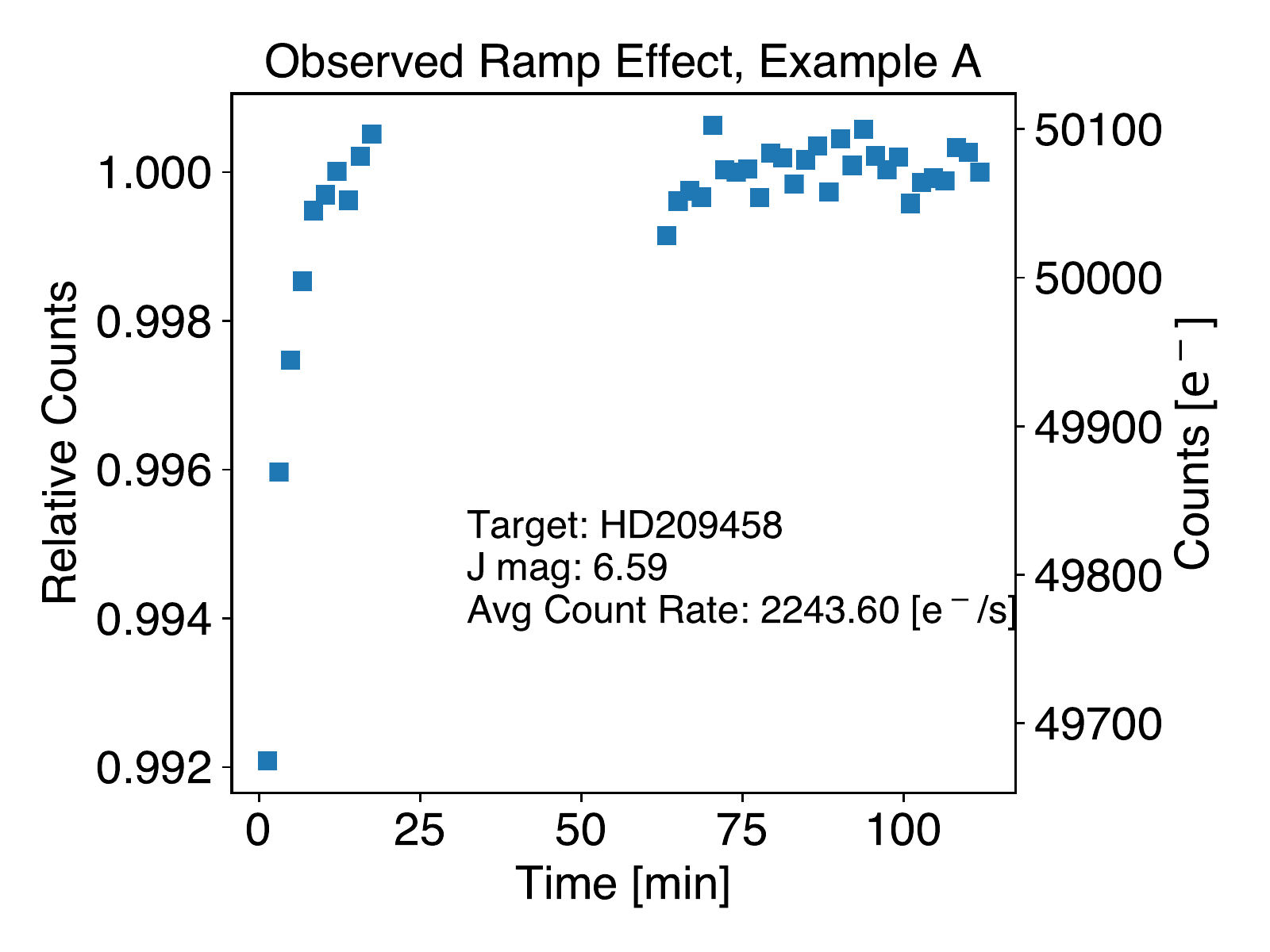}{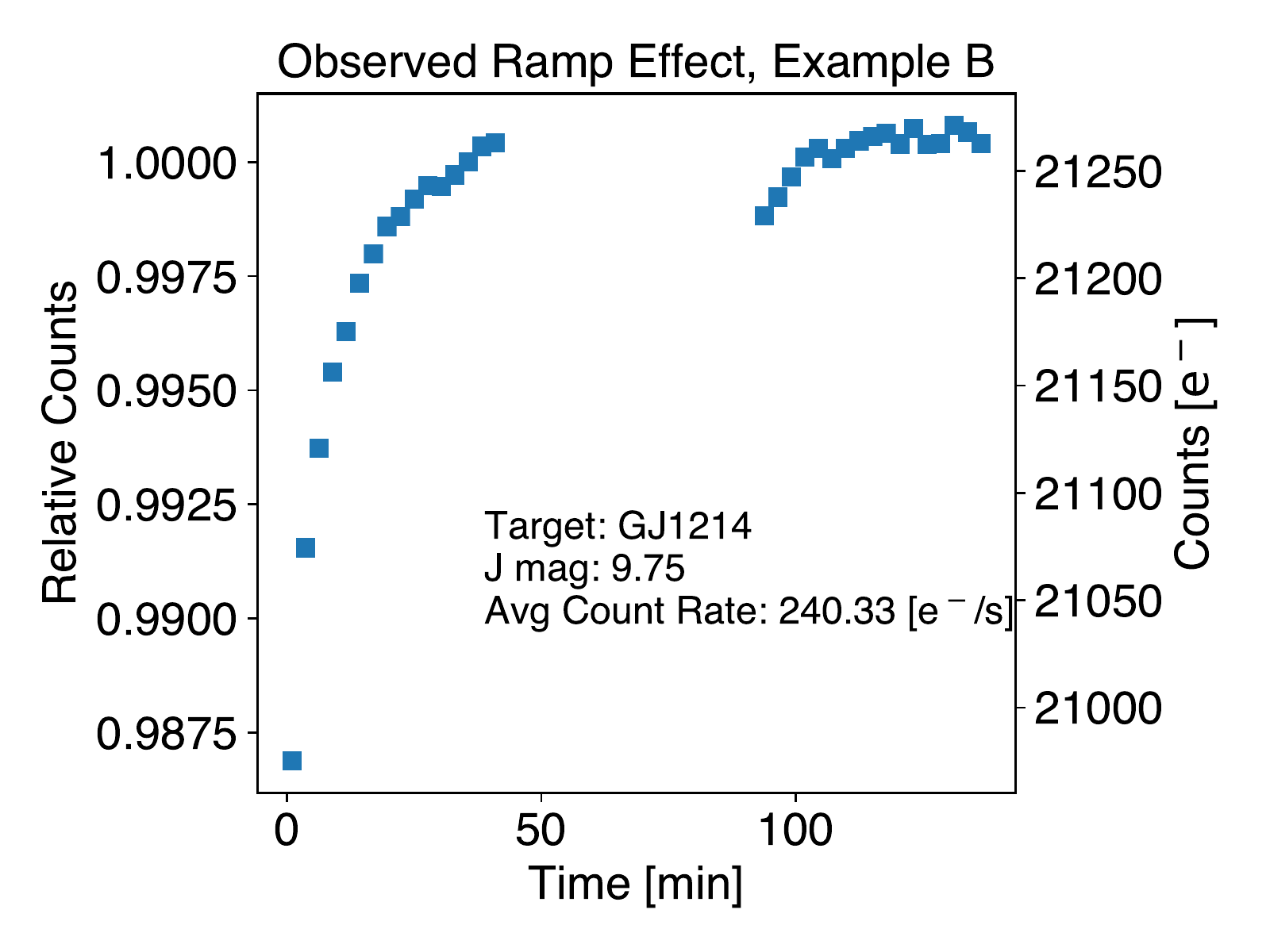}
  \plottwo{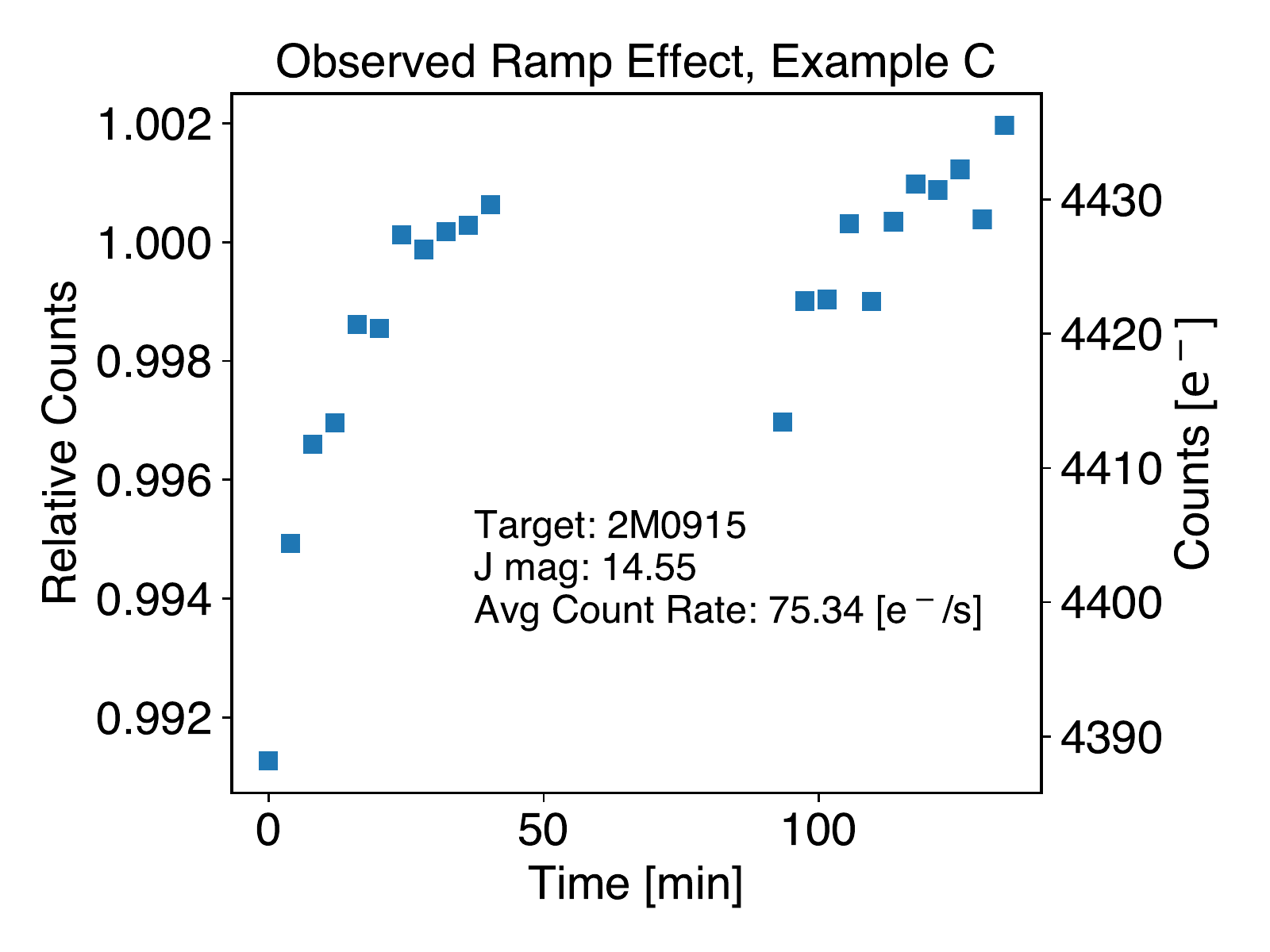}{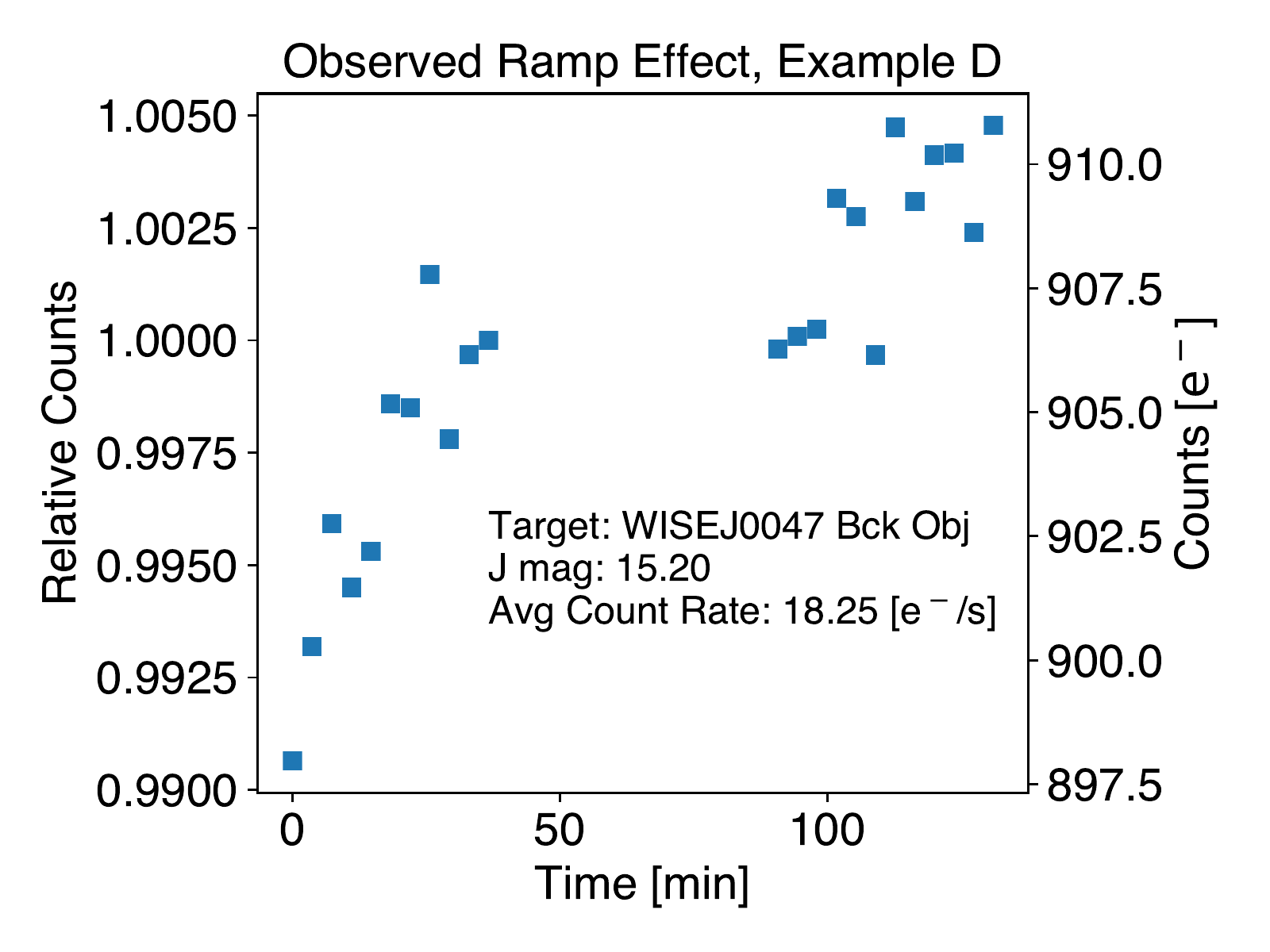}
  \caption{Different manifestations of the ramp effect are presented in various observations. With
    different target brightness, exposure times, and exposure
    sequences, the ramps have different shapes.}
  \label{fig:obsRamp}
\end{figure*}

The Hubble Space Telescope Wide Field Camera 3 (\HST\WFC) \glenn{near
  infrared (near-IR) channel}, is one of the most powerful and most
frequently used instruments for exoplanet atmosphere
observations. With its exceptional sensitivity and high photometric
stability, \WFC plays an essential role in high-cadence time domain
observations, including transit grism spectroscopy \citep[e.g.,
][]{Swain2013, Kreidberg2014} and rotational phase mapping
\citep[e.g.,][]{Buenzli2012,Apai2013,Lew2016,Zhou2016}. However,
instrumental systematics in its near-IR channel either related to the
detector \citep[for a summary see][]{Wakeford2016} or the grism
geometry \citep[for a detailed description and correction,
see][]{Varley2015} prevent the attainment of photon noise limited
performance. The often-called ``\textit{ramp effect}''
\citep[e.g.,][]{Berta2012}, is an approximately exponentially-shaped
signal in the time domain. It is the most significant systematic
affecting photometric efficacy. Different empirical profiles of the
ramp effect were identified in various HST time-resolved observation
datasets (Figure \ref{fig:obsRamp}).
It is expected that other instruments that employ detectors with similar
architectures, such as JWST/NIRCAM and NIRSPEC, may also suffer from similar systematics. 

Great effort has been devoted to the calibration of the ramp
effect. Until now, the most popular and successful method has been the
``divide-out-of-transit'' method \citep[e.g.,][]{Berta2012}. This
method uses an exponential/polynomial function (empirically derived
from the out-of-transit parts of the light curves) to correct the ramp
effect. Several of the most precise \HST transit spectroscopic
observations \citep[e.g.,][]{Berta2012, Kreidberg2014} have adopted
this correction. However, this method has three important limitations.
First, the ramp effect is significantly more severe in the first HST
orbit than in the rest of the observations in sequential target
visibility periods,\footnote{During each orbit of HST ($\sim93$
  minutes) a typical target is visible for 40--54 minutes, depending on
  its celestial position. Therefore, HST only acquires science
  exposures during a fraction of each period, i.e., in the “visibility
  period}, and the first orbit cannot be
corrected by the divide-out-of-transit method.  Although the
difference in systematics between the different orbits is not well
understood, it is widely assumed that the telescope needs to thermally
settle, i.e., reach some sort of equilibrium. Therefore, the data from
the first orbit of each visit are always excluded from data analysis,
which results in the loss of approximately 100--200 orbits up to
\glenn{the end of \HST cycle 24}. Second, the out-of-transit method
requires a flat photometric baseline to fit the empirical functions
that are used for correction. When intrinsic variability is present,
such as in hot-Jupiter orbital phase curves and in rotational phase
mapping observations of brown dwarfs and directly imaged exoplanets,
this method is not reliable. Third, this method is not based on an
understanding of the underlying physical mechanisms of the
systematics.
The ad hoc nature of this method makes it difficult to evaluate the
applicability of this instrumental systematic when observations with
different detector readout modes and illumination levels are obtained
with different instrument configurations.
A physically motivated ramp-effect model that enables the retention of
all photometric data with post-processing (ramp calibration) is
therefore highly desired.


Charge trapping has been suggested to be the major cause of the ramp
effect for infrared detectors. \citet{Agol2010} used electron trapping
to attempt to explain the ramp effect of \textit{Spitzer}'s IRAC
detectors, and they empirically described this effect with two
exponential functions. Furthermore, in \WFC data the amplitudes of the ramp
were found to be related to the exposure count levels
\citep[e.g.,][]{Berta2012}, a finding that is broadly consistent with
charge trapping.

Instead of using empirically estimated exponential functions to
calibrate electron trapping, in this paper, we developed a physical
model RECTE (Ramp Effect Charge Trapping Eliminator), which is described by
the numbers of charge carrier traps, the trapping efficiency and the trap
lifetimes in every pixel on the detector. We show that this model
works extremely well with time-resolved observations taken in both
scanning and staring modes. We explain the details of the model (\S
\ref{sec:model}), describe the model results (\S \ref{sec:results}), demonstrate two examples of model applications (\S \ref{sec:app}), and
discuss the correction results, future observation planning
suggestions, and model extensions for future instruments (\S \ref{sec:discussion})
in the rest of the paper.

\section{The Ramp Effect Model}\label{sec:model}

\subsection{Physical Background}

The \WFC infrared channel detector is a $\mbox{1k}\times\mbox{1k}$
HgCdTe array \citep{Dressel2016}. The detector was manufactured by
Teledyne Scientific \& Imaging, and has an architecture of hybrid
HgCdTe device grown on a CdZnTe substrate and indium-bonded to a
Hawaii-1R MUX \citep{Baggett2008}.  The basic elements of the detector
array are photodiodes.  Photons are absorbed by the diodes to produce
free charge carriers. The free charge carriers travel through the
diodes following the electric potential and reach the depletion
region of the diodes. The electric field maintained by the contact
potential drives the charge carriers across the P-N junction, and the
collected charge carries change the circuit gate voltage that are
measured as the signal \citep{Rieke2012}.

A highly repeatable characteristic of this type of detector is that
the P-N junction is de-biased as the signal is accumulated. This can
change the response of the detector during a series of exposures
\citep{Rieke2012}, and is suggested to be the primary source of image
persistence \citep{Smith2008a}. The latter authors proposed that
charge traps in the depletion region of the P-N junction are
responsible for the image persistence, and based on this assumption
established a qualitative theoretical framework to explain the image
persistence phenomenon.
Ibid, charge carriers can be trapped as they diffuse across the
depletion region. The trapping of the charge carriers lowers the level
of the signal measured in the individual exposures. At the end of
the exposure, the original width of the depletion  is restored by reset, but
the trapped charge carriers remain in the depletion region. In
subsequent frames, the trapped charge carriers are released and
diffuse back to the undepleted region, which generates a signal. This
signal is manifested either as image persistence in the dark frame, or
as a ramp-shaped light curve in exposure series.

\citet{Smith2008b} applied this model to calibrate the image
persistence in the HgCdTe array of the SuperNova Acceleration
Probe. They found that the integrated persistence profiles can
  be calibrated using a double-exponential profile, of which the time
  constants of two exponential functions differed by more than an
  order of magnitude. They suggested that the two trap populations
could be the result of the two types of charge carriers, electrons and
holes, behaving differently.  Furthermore, they found that the
trapping efficiency has relatively small spatial variation, which
provided evidence that individual pixels had similar trapping
properties. Studies \citep[e.g.,][]{Anderson,Long2015a} explored to
study the charge carrier trap properties of the \WFC near-IR detector
and detectors that share similar physical
architectures. \citet{Long2015b} created a model for the persistence
of \WFC near-IR detector. These studies focused on the persistence
behavior of the detector with a timescale of several minutes after
high count level exposures. These studies aimed to describe the {\em
  effects} of charge trapping on the data by fitting empirical
functions to the observed trends, an approach that only works
imperfectly and only applicable within sets of self-similar
observations. The model we introduce here invokes the {\em
    same} physically-motivated conceptual picture as previous
  studies\citep[e.g.,][]{Smith2008a}; however, our approach differs
  from the past qualitative and quantitative models in two important
  aspects:

  First, our model assumes that charge trapping occurs immediately
  after the photons are absorbed and not later, i.e., charge trapping
  occurs only when the detector is illuminated. This difference
  results in a different behavior of the detector, which we
  demonstrate is more consistent with the existing observations than
  the predictions of the traditional model that assumes continuing
  trapping even in the non-illuminated state of the detector.

Second, our model aims at correcting for the ramp effect and {\em not}
the image persistence. The later effect is only important at very high fluence levels. Therefore, our model does not aim to incorporate
a complete description of charge trapping and release mechanisms
and the effects of nonlinearity in the extreme case of
near-saturation levels. Comparisons of our model predictions to
observations shows that our model reproduces very well datasets taken
at moderately high fluence levels but not those taken at close to
saturation levels. This indicates that physical effects not considered in
our model play important roles when the detector is driven to
saturation.

\subsection{Model Descriptions}
\newcommand{\ntrapo}{\ensuremath{E_\mathrm{tot}}\xspace}
\newcommand{\flux}{\ensuremath{f}}
\newcommand{\dd}{\mathrm{d}\xspace}
\newcommand{\trapped}{\ensuremath{E}\xspace}
\newcommand{\dtrap}{\ensuremath{\Delta \trapped}\xspace}

\begin{figure*}[!t]
  \centering
  \includegraphics[width=0.33\textwidth]{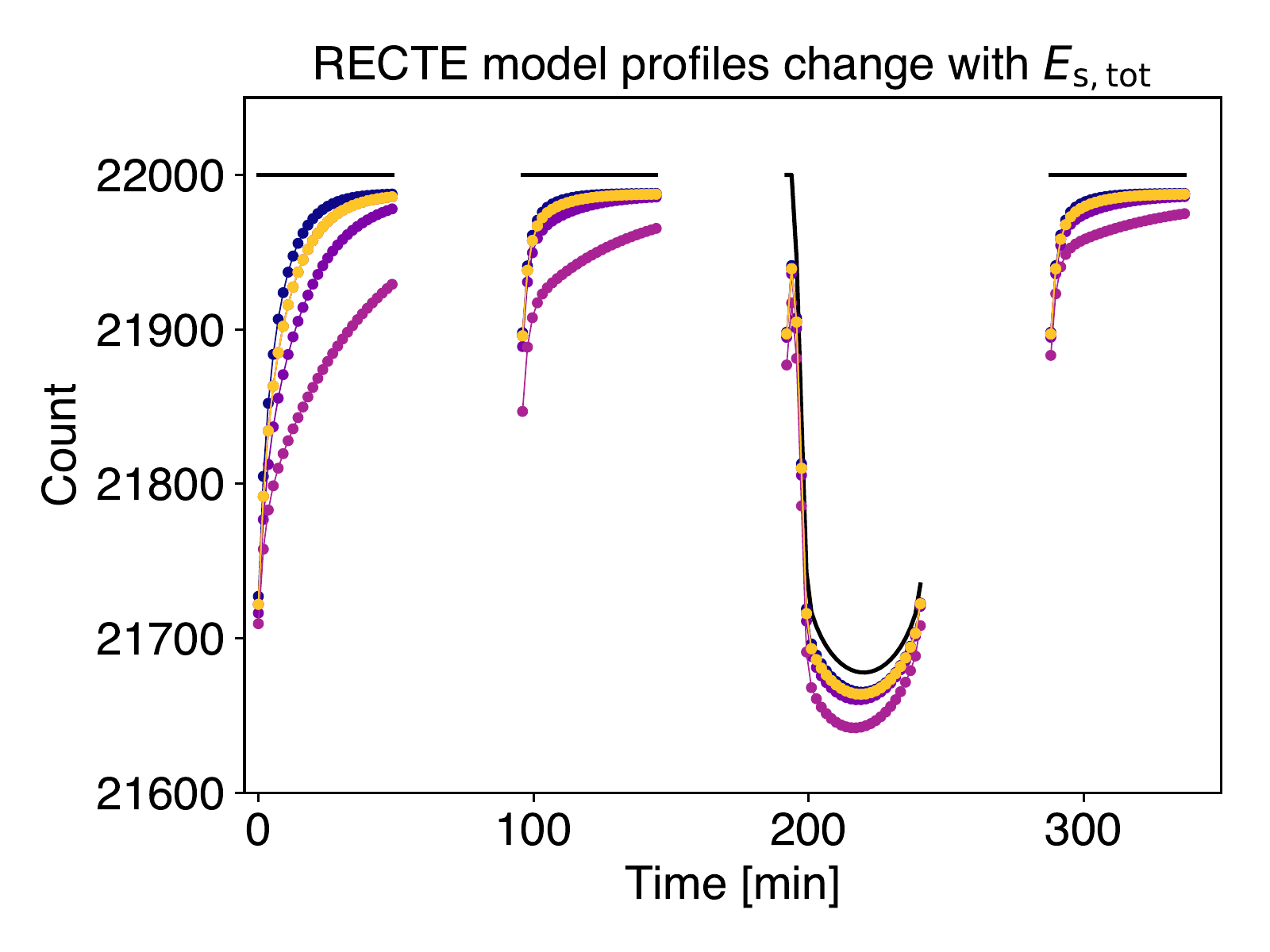}
  \includegraphics[width=0.33\textwidth]{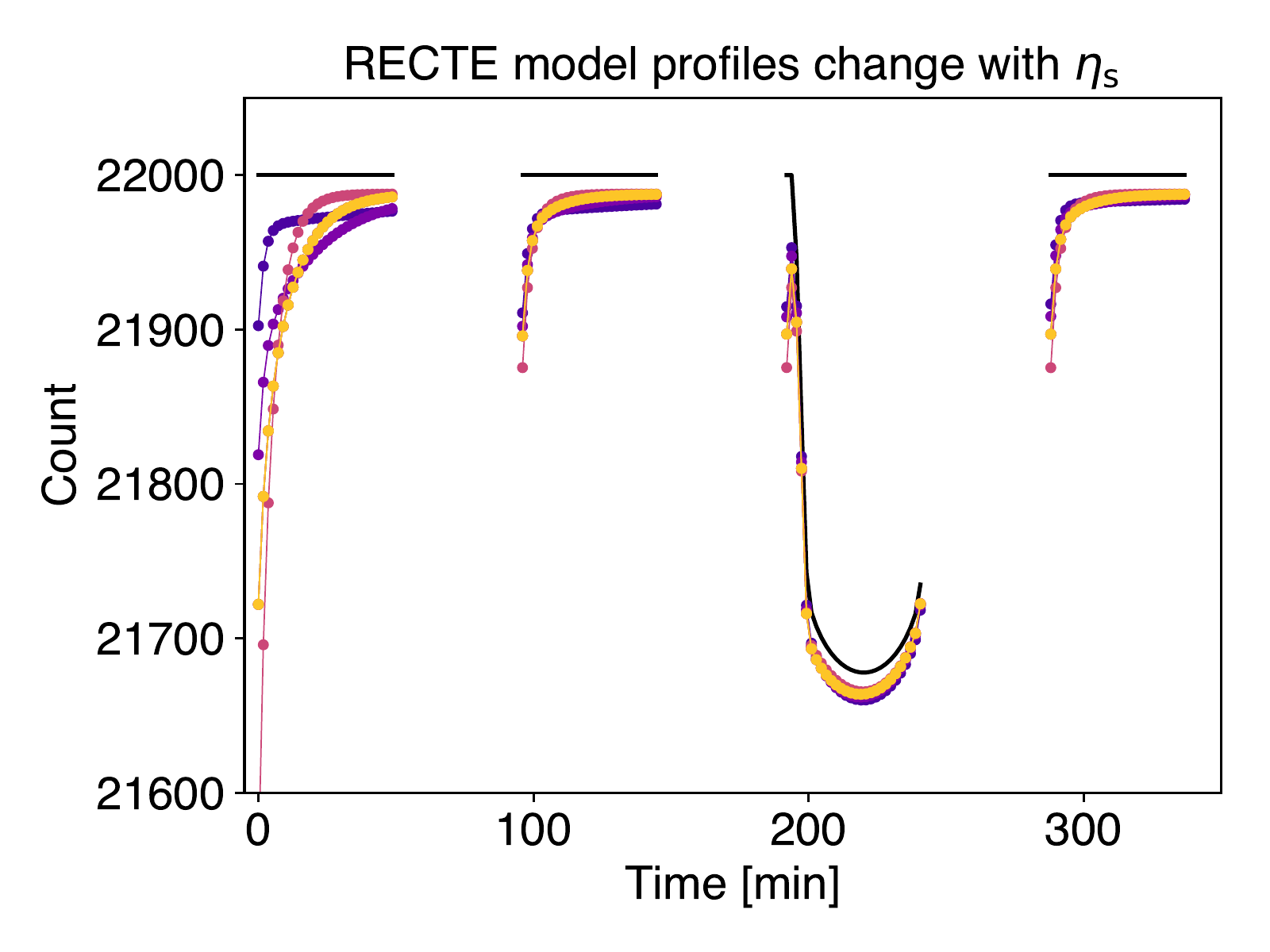}
  \includegraphics[width=0.33\textwidth]{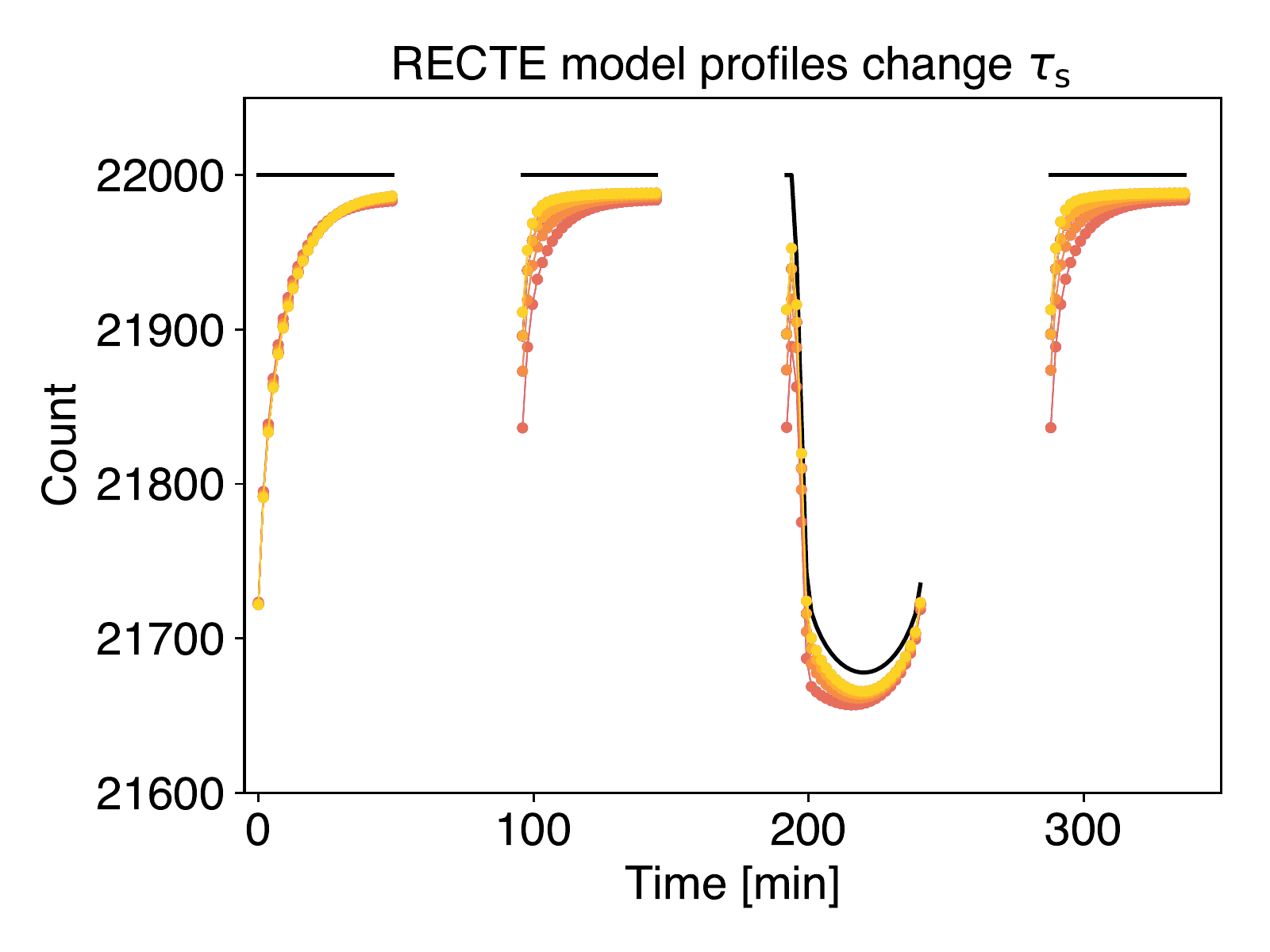}\\
  \includegraphics[width=0.33\textwidth]{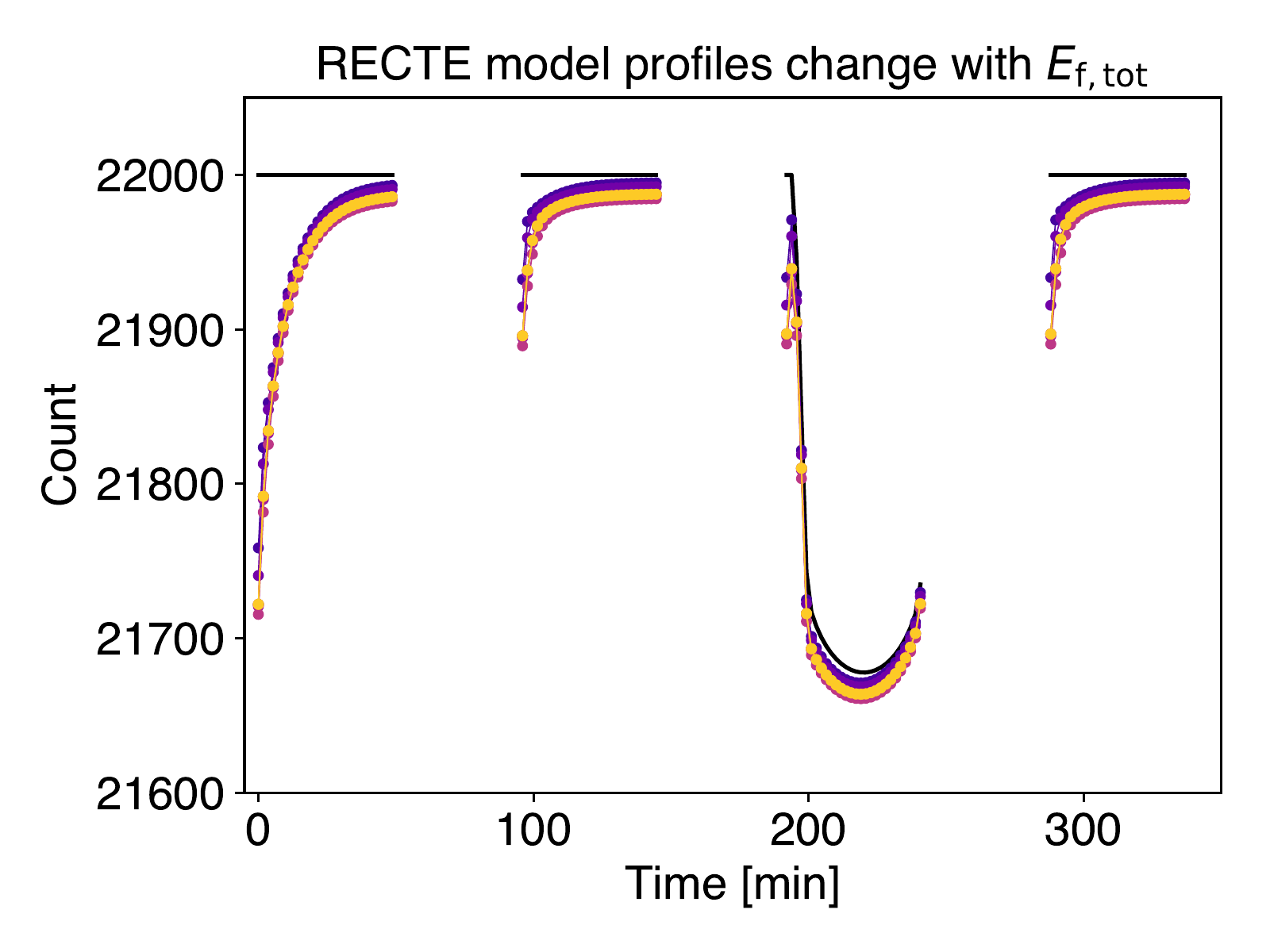}
  \includegraphics[width=0.33\textwidth]{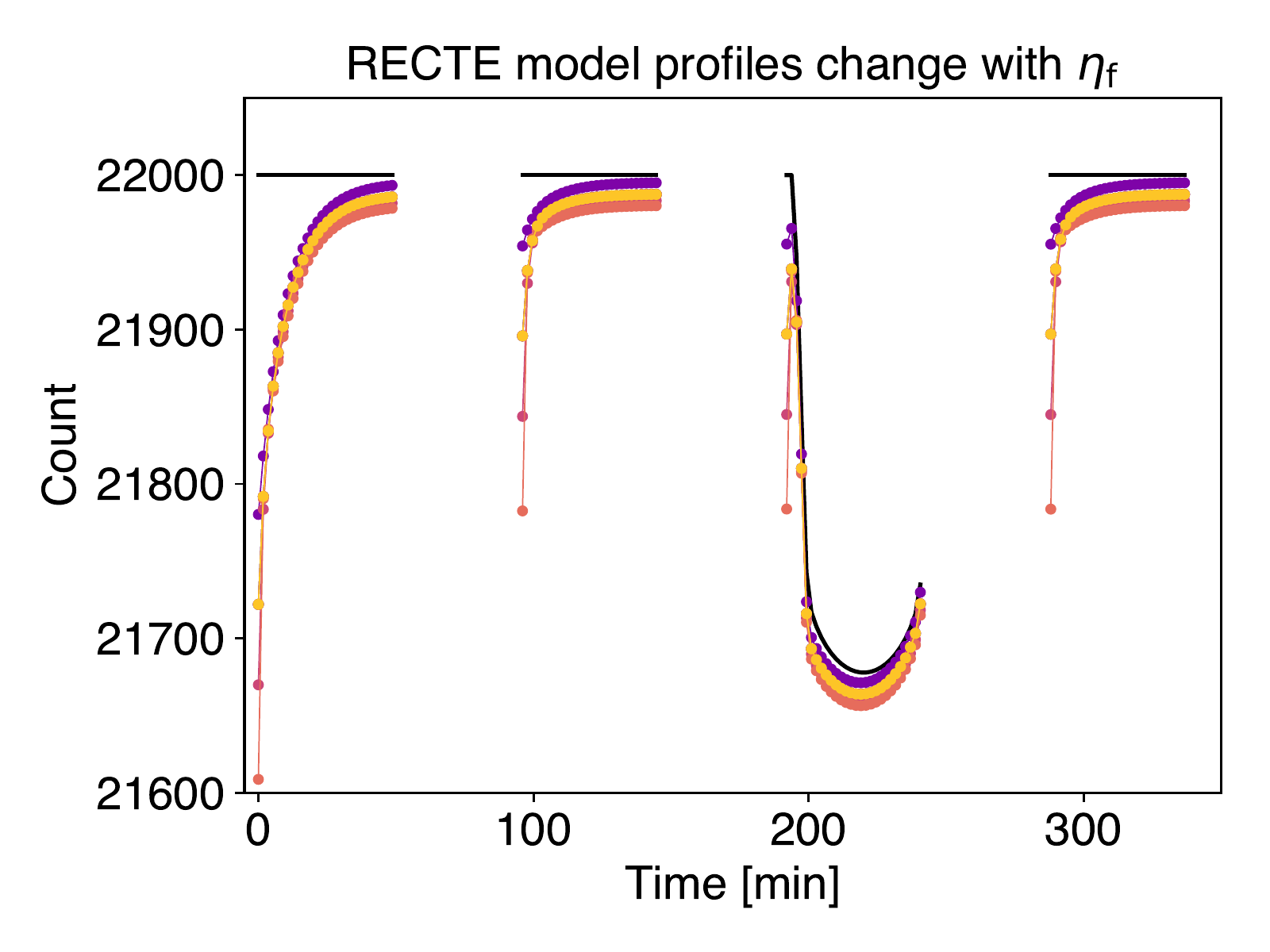}
  \includegraphics[width=0.33\textwidth]{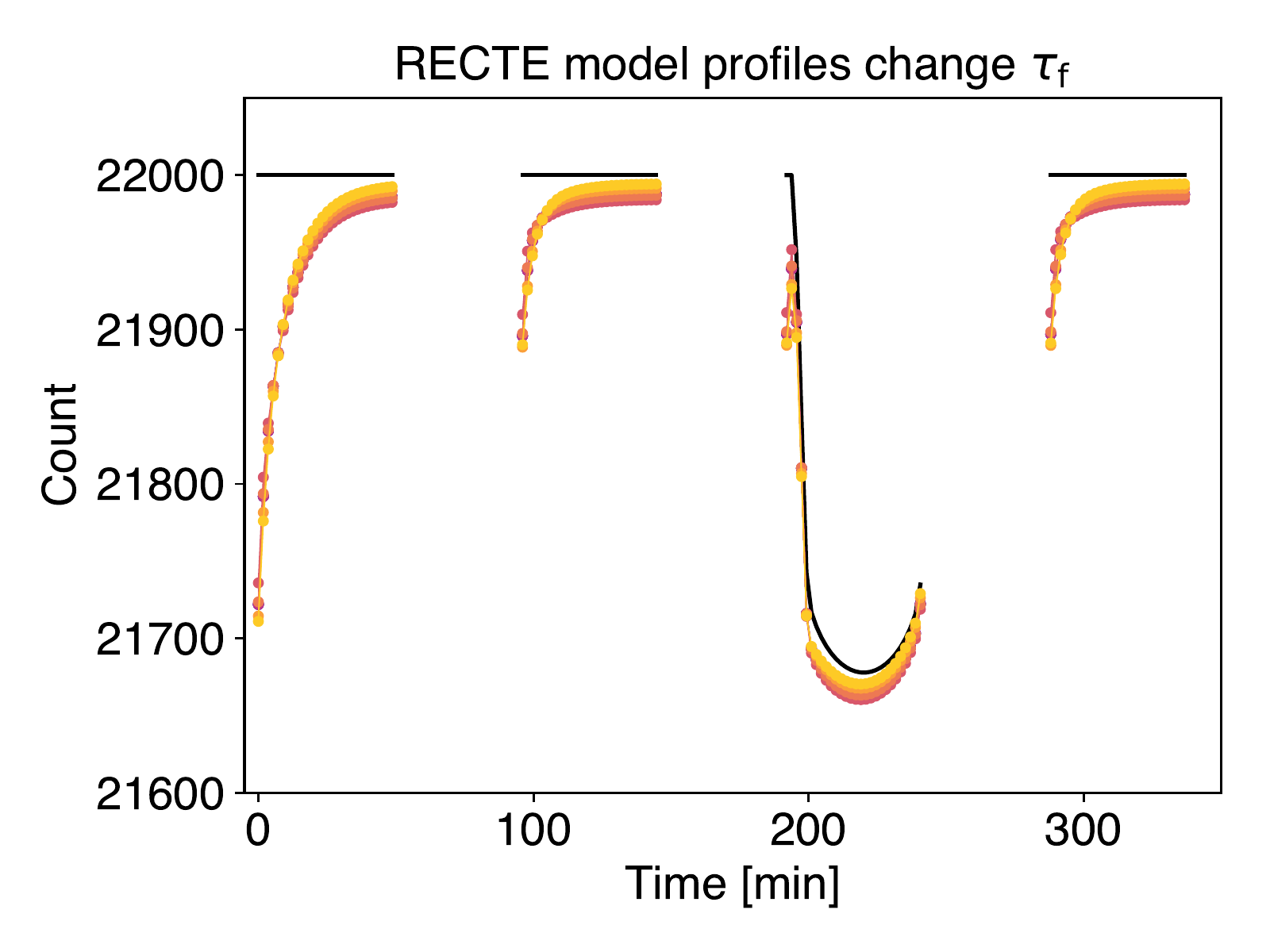}
  \caption{Four orbit transit observations of GJ1214b with SPARS10,
    NSAMP=11 time series simulated by RECTE with different parameter
    combinations. By adjusting \ntrapo (\textit{left}), $\eta$
    (\textit{middle}), or $\tau$ (\textit{right}), the shapes of the
    ramps change differently for a fixed constant incoming flux. The
    lines are color coded in the way that the darker the line, the
    larger the changing model parameter is. The black lines are the
    light curves with no ramp effect.}
  \label{fig:model}
\end{figure*}

Our model is based on the charge carrier trapping theory of
\citet{Smith2008a}. However, instead of fitting empirically derived
exponential functions, our model enables us to quantitatively model
the charge carrier trapping processes, so that we can precisely
calibrate ramp-effect-impacted time-resolved observations made
with \WFC. In order to quantitatively constrain the behavior of \WFC
detector, we generalize the theory with the following assumptions.
\begin{enumerate}
\item The detector pixels have two populations of charge carrier
  traps: a slow trap population that releases trapped particles with a
  long trapping lifetime and a fast trap population that releases
  trapped particles with a short trapping lifetime.  The total numbers
  of traps per pixel for the two population are $E_{\mathrm{s, tot}}$
  and $E_{\mathrm{f, tot}}$. The power-law persistence trend
    seen in studies of \citet{Long2012,Long2015a} suggests the
    possibility of a more complex nature of the traps. We focus on two
    trap population model in this work and reserve the possibility to
    extend the model to multiple trap populations, which may become
    important when the detector is illuminated at levels close to
    saturation.
  
  \item Charge carriers stimulated by incoming photon fluxes can fill
    the two populations of traps with efficiencies of
    $\eta_{\mathrm{s}}$ and $\eta_{\mathrm{f}}$. \glenn{The states of
      the two populations' traps are independent.} The numbers of the
    trapped charge carriers at time $t$ are $\trapped_{\mathrm{s}}(t)$
    and $\trapped_{\mathrm{f}}(t)$. The charge carrier trapping rates
    are proportional to incoming flux \flux(t) [$\mathrm{e^-/s}$] and
    to the number of unfilled traps
    ($\bar{\trapped}(t) = \ntrapo - \trapped(t)$). With these
    assumptions the charge carrier trapping rate can be expressed as
    Equation \eqref{eq:gain}.  \footnote{No subscripts of
      $_{\mathrm{s}}$ and $_{\mathrm{f}}$ means the expressions work
      for both trap populations} \footnote{Superscription $^{+}$
      denotes trap gaining processes, and $^{-}$ denotes trap
      releasing processes.}
  \begin{equation}
    \frac{\dd \trapped(t)^+}{\dd t} = \eta\,\flux(t)\,\frac{\ntrapo-\trapped(t)}{\ntrapo}\label{eq:gain}
  \end{equation}
  
\item The two populations of trapped charge particles have lifetimes
  of $\tau_{s}$
  and $\tau_{f}$.  Under no illumination, the
  number of trapped charges follows exponential decay:
  \begin{align}
    &E(t) = E(t_{0})\exp\Bigl(-\dfrac{t-t_{0}}{\tau}\Bigr) \quad \mbox{when $\flux=0$}\\
    &\frac{\dd \trapped(t)^-}{\dd t} = - \frac{\trapped(t)}{\tau}\label{eq:lose}
  \end{align}.
  Combining equation \eqref{eq:gain} and \eqref{eq:lose}, the complete form for trap number change follows:
  \begin{equation}
    \label{eq:1}
    \frac{\dd \trapped(t)}{\dd t} = \eta\,\flux(t)\,\frac{\ntrapo-\trapped(t)}{\ntrapo} - \frac{\trapped(t)}{\tau}
  \end{equation}


 
\item As traps are filled during the exposures, the charge carrier
  trapping rates decrease. Therefore, the number of detected electrons
  per unit time increases, giving the characteristic ``ramp''
  shape. When all traps are filled, or the charge carrier release rate
  is equal to the trapping rate, the detected flux will be equal to
  the true flux. In addition, a notable number of charge carriers are
  released \glenn{between the target visibility periods, when the
    detector is not illuminated by the target.} Therefore, the light
  curves usually rise up again at the beginning of each subsequent
  orbit (see Figure \ref{fig:obsRamp}).

\item The number of occupied traps is not necessarily zero at the
  beginning of each visit because 1) previous observations may have
  illuminated the same pixels as current exposures; 2) the detector
  received flux before the first exposure, as \WFC has no shutter for
  the detector, and is occassionally unintentionally illuminated
  during telescope slewing. To reflect these possibilities we
  introduced parameters $E_{\mathrm{s}, 0}$ and $E_{\mathrm{f}, 0}$
  that represent the initially occupied number of traps. These
  parameters can vary from visit to visit, as well as from pixel to
  pixel.
\item Furthermore, because \WFC may be illuminated by the target before the
  beginning of the science exposure series, parameters
  $\dtrap_\mathrm{s}$ $\dtrap_\mathrm{f}$ are included to represent
  extra trapped charges during the time between visibility periods.
\item After evaluating a number of different data sets we found no
  evidence for trapping parameters ($\ntrapo$, $\eta$, and $\tau$) to
  change with time; therefore, we assume these parameters are intrinsic
  to the detector and constant with time. $E_{0}$ and $\dtrap$ can be
  different for different observations.
\end{enumerate}

Table \ref{tab:paramlist} summarizes the key parameters of our model and their physical meaning.

\floattable
\begin{deluxetable}{cll}
\tablecaption{Parameter list \label{tab:paramlist}\authorcomment1{column unit added}}
\tablecolumns{3}
\tablenum{1}
\tablewidth{0pt}
\tablehead{
  \colhead{Parameter} &
  \colhead{Unit} &
  \colhead{Description}
}
\startdata
$E_{\mathrm{s, tot}}$, $E_{\mathrm{f, tot}}$ & count &Total numbers of slow and fast traps\\
$\eta_{\mathrm{s}}$, $\eta_{\mathrm{f}}$ & - &Trapping efficiencies for slow and fast traps\\
$\tau_{\mathrm{s}}$, $\tau_{\mathrm{f}}$ & second &Trap lifetimes for slow and fast traps\\
$E_{\mathrm{s},0}$, $E_{\mathrm{f},0}$ & count &Number of slow and fast traps occupied at the beginning of  visit.\\
$\dtrap_{\mathrm{s}}$, $\dtrap_{\mathrm{f}}$ & count & Extra number of charge carriers captured by slow and fast traps during the observation gaps.\\
$f$ & count/s & illumination flux\\
\enddata
\end{deluxetable}

The model is presented as a flowchart in Figure \ref{fig:flowchart}:

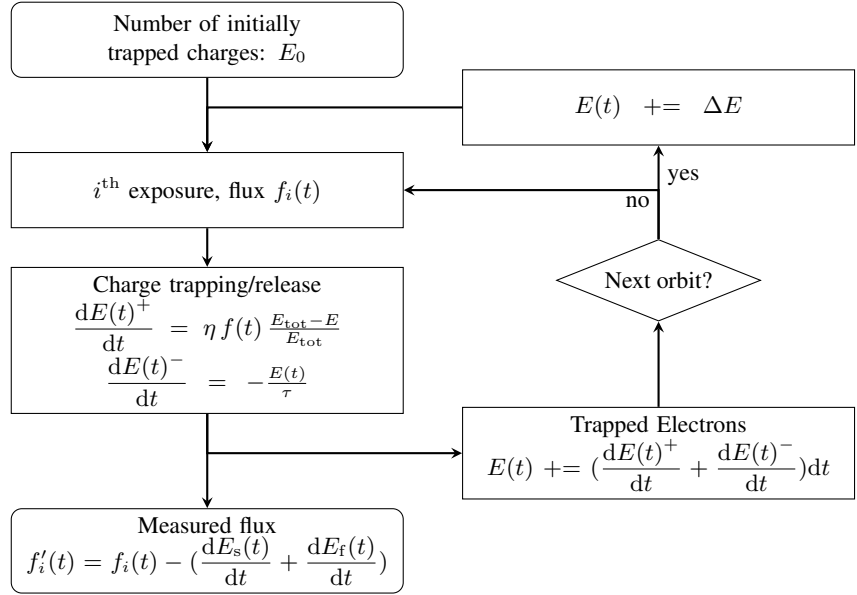
\begin{figure*}
  \tikzstyle{startstop} = [rectangle, rounded corners, minimum width=2cm, minimum height=1cm,text centered, draw=black, text width=5cm]
\tikzstyle{process} = [rectangle, minimum width=2cm, minimum height=1cm, text centered, draw=black, text width=5cm]
\tikzstyle{decision} = [diamond, minimum width=2cm,  text centered, draw=black, aspect=2.5]
\tikzstyle{arrow} = [thick,->,>=stealth]

\begin{center}
\begin{tikzpicture}[node distance=2cm]
  \node (start) [startstop] {Number of initially trapped charges: $E_0$ };
  \node (exposure) [process, below of=start] {$i^{\mathrm{th}}$ exposure, flux $\flux_i(t)$};
  \node (trap) [process, below of=exposure] {Charge trapping/release\\ $\dfrac{\dd \trapped(t)^+}{\dd t} = \eta\,\flux(t)\,\frac{\ntrapo-\trapped}{\ntrapo}$ $\dfrac{\dd \trapped(t)^-}{\dd t} = - \frac{\trapped(t)}{\tau}$};
  \node (measurement) [startstop, below of=trap, yshift=-0.8cm] {Measured flux\\ $\flux'_i(t) = \flux_i(t) - (\dfrac{\dd \trapped_\mathrm{s}(t)}{\dd t} + \dfrac{\dd \trapped_\mathrm{f}(t)}{\dd t})$};
  \node (nTrapped) [process, right of=trap, xshift=4cm,
  yshift=-1.5cm] {Trapped Electrons\\ $E(t) \mathrel{+}= (\dfrac{\dd
      \trapped(t)^+}{\dd t} + \dfrac{\dd \trapped(t)^-}{\dd t}) \dd t$};
  \node (changeorbit) [decision, above of=nTrapped, yshift=0.3cm] {Next orbit?};
  \node (extraE) [process, above of=changeorbit, yshift=0.3cm] {$E(t) \mathrel{+}= \dtrap$};
  \draw [arrow] (start) -- (exposure);
  \draw [arrow] (exposure) -- (trap);
  \draw [arrow] (trap) -- (measurement);
  \draw [arrow] (trap) |- (nTrapped);
  \draw [arrow] (nTrapped) -- (changeorbit);
  \draw [arrow] (changeorbit) -- node [anchor=west,yshift=.2cm]{yes} (extraE);
  \draw [arrow] (changeorbit) |- node [anchor=east,yshift=-.15cm]{no} (exposure);
  \draw [arrow] (extraE) -| (exposure);
\end{tikzpicture}
\end{center}
  \caption{RECTE model processes presented with a flowchart.}
  \label{fig:flowchart}
\end{figure*}

\subsection{Mathematical solutions for ramp profiles}
From equation \eqref{eq:gain} and \eqref{eq:lose}, we can express $E(t)$ as a differential equation:
\newcommand{\ddfrac}[2]{\ensuremath{\frac{\dd#1}{\dd#2}}}
\begin{equation}
  \label{eq:diff}
  \ddfrac{E}{t} = \ddfrac{E^+}{t} + \ddfrac{E^-}{t} = \eta f(t) - \left(\frac{\eta f(t)}{\ntrapo} + \frac{1}{\tau}\right)E
\end{equation}
The formal solution of \eqref{eq:diff} is
\begin{equation}
  \label{eq:2}
    E(t) = \left[\int \eta f(t) \mathrm{e}^{\int \left(\frac{\eta f(t)}{\ntrapo} + \frac{1}{\tau}\right) \dd t} \dd t+C \right]
    \mathrm{e}^{-\int\left(\frac{\eta f(t)}{\ntrapo} + \frac{1}{\tau}\right) \dd t}
  \end{equation}

We note that $E_\mathrm{s}(t)$ and $E_\mathrm{f}(t)$ are calculated separately using equation \eqref{eq:2}, as we assume the states of the two populations are independent.

Therefore, the measured flux is
\begin{equation}
  \label{eq:3}
  f'(t) = f(t) - \ddfrac{E_\mathrm{s}(t)}{t} - \ddfrac{E_\mathrm{f}(t)}{t}
\end{equation}

If the incoming photon flux is constant, equation \eqref{eq:diff}
has an analytical solution of

\begin{equation}
  \label{eq:4}
  E(t) = \frac{\eta f}{\frac{\eta f}{\ntrapo} + \frac{1}{\tau}} + \left(E_0 - \frac{\eta f}{\frac{\eta f}{\ntrapo} + \frac{1}{\tau}}\right) \mathrm{e}^{-\left(\frac{\eta f}{\ntrapo} + \frac{1}{\tau}\right)t}
\end{equation}

where $E(t=0)=E_0$. The observed flux is the intrinsic flux subtracted
by the charge carriers that are trapped:
\begin{equation}
  \label{eq:5}
  \begin{split}
    f'(t) = f &- \ddfrac{E_\mathrm{s}(t)}{t} - \ddfrac{E_\mathrm{f}(t)}{t}\\
    = f &- \left[\eta_\mathrm{s} f - E_\mathrm{s,0}(t)\left(\frac{\eta_\mathrm{s} f}{E_\mathrm{s,tot}} + \frac{1}{\tau_\mathrm{s}}\right)\right]\mathrm{e}^{-\left(\frac{\eta_\mathrm{s} f}{E_\mathrm{s,tot}} + \frac{1}{\tau_\mathrm{s}}\right)t}\\
    &- \left[\eta_\mathrm{f} f - E_\mathrm{f,0}(t)\left(\frac{\eta_\mathrm{f} f}{E_\mathrm{f,tot}} + \frac{1}{\tau_\mathrm{f}}\right)\right]\mathrm{e}^{-\left(\frac{\eta_\mathrm{f} f}{E_\mathrm{f,tot}} + \frac{1}{\tau_\mathrm{f}}\right)t}
    \end{split}
\end{equation}

In the following, we discuss a few manifestations of our model.
\begin{enumerate}
\item At the beginning of the observations and in the case where $E_0$
  is close to zero, the measured flux is $f'(t) =
  (1-\eta_\mathrm{s}-\eta_\mathrm{f})f$. Therefore, the amplitude of the ramp is mainly
  controlled by $\eta_\mathrm{s} + \eta_\mathrm{f}$.

\item  When the number of filled traps reaches
\begin{equation}
  \label{eq:6}
  E = \frac{\eta f}{\frac{\eta f}{\ntrapo} + \frac{1}{\tau}}
\end{equation}
where  released electrons and trapped electrons are in
equilibrium. In this case, a constant incoming flux results in a flat measured light curve.
\item If the irradiated count rates are low, $\left(\frac{\eta f}{\ntrapo} + \frac{1}{\tau}\right) \ll 1$, the measured flux will be
  \begin{equation}
    \label{eq:7}
    \begin{split}
      f'(t) =& (1 - \eta_\mathrm{s} - \eta_\mathrm{f})f+\\
      & \left(\frac{\eta_\mathrm{s}^2 f^2}{E_\mathrm{s, tot}} + \frac{\eta_\mathrm{s}f}{\tau_\mathrm{s}} + \frac{\eta_\mathrm{f}^2 f^2}{E_\mathrm{f, tot}} + \frac{\eta_\mathrm{f}f}{\tau_\mathrm{f}}\right)t
      \end{split}
  \end{equation}
  In this case, the ramp will become a linear profile.
\end{enumerate}


\subsection{Constraining the model parameters}

We constrain our model by primarily using the scanning mode observations of
the transiting exoplanet GJ1214b \citep[HST program 13021]{Kreidberg2014}. This dataset
includes 15 visits that were taken between September 2012 and
August 2013.  Each of the 15 visits included 4 orbits G141 transiting
spectroscopic observations to observe one transit of GJ1214b, and the
transits all occurred within the third orbit of each visit observations. The first 5
visits were taken with one-directional scanning and SPARS10, NSAMP=13
exposures, and the last 10 visits are taken with two-directional (round
trip) scanning and SPARS10, NSAMP=15 exposures.

We used the first two-orbit observations of every visit to constrain
the model parameters, because the first two orbits had intrinsically
flat light curves. The last two orbits in each visit were used to test
the validity of the model. \citet{Kreidberg2014} discarded three
visits for their analysis because of star-spot crossing or guiding
failure. The first two orbits of these visits were not affected by
these problems. Therefore, we used all 15 visits to constrain the
model parameters, but discarded the same three visits for model
testing and validation.

\begin{deluxetable*}{llcllll}
\tablecaption{\WFC data used for model fitting and testing \label{tab:params}}
\tablecolumns{4}
\tablenum{2}
\tablewidth{0pt}
\tablehead{
\colhead{ID} &
\colhead{PI}&
\colhead{No. of orbits}&
\colhead{Subframe}&
\colhead{Exp. mode}&
\colhead{Peak count levels [e$^{-}$]}&
\colhead{Observing Modes}
}
\startdata
12181&Deming&8&512&RAPID, NSAMP=16&$6.4\times10^{4}$&staring\\
12181&Deming&4&256&SPARS10, NSAMP=5&$4.9\times10^{4}$&one direction scanning\\
12251&Berta-Thompson&12&512&RAPID, NSAMP=7&$6.4\times10^4$&staring\\
12314&Apai&6&256&SPARS25, NSAMP=12&$1.9\times10^4$&staring\\
13021&Bean&60&256&SPARS10, NSAMP=13,15&$2.3\times10^{4}$&one/two
direction scanning\\
13573&Long&8&256&SPARS10, NSAMP=5,6&$2.8\times
10^4$/$3.8\times10^4$&two direction scanning\\
14241&Apai& 28 &256 &SPARS10, NSAMP=5-8&$1.2\times 10^4 $- $3.0\times10^4$& staring\\
\enddata
\end{deluxetable*}

\begin{figure}[!h]
  \centering
  \includegraphics[width=\columnwidth]{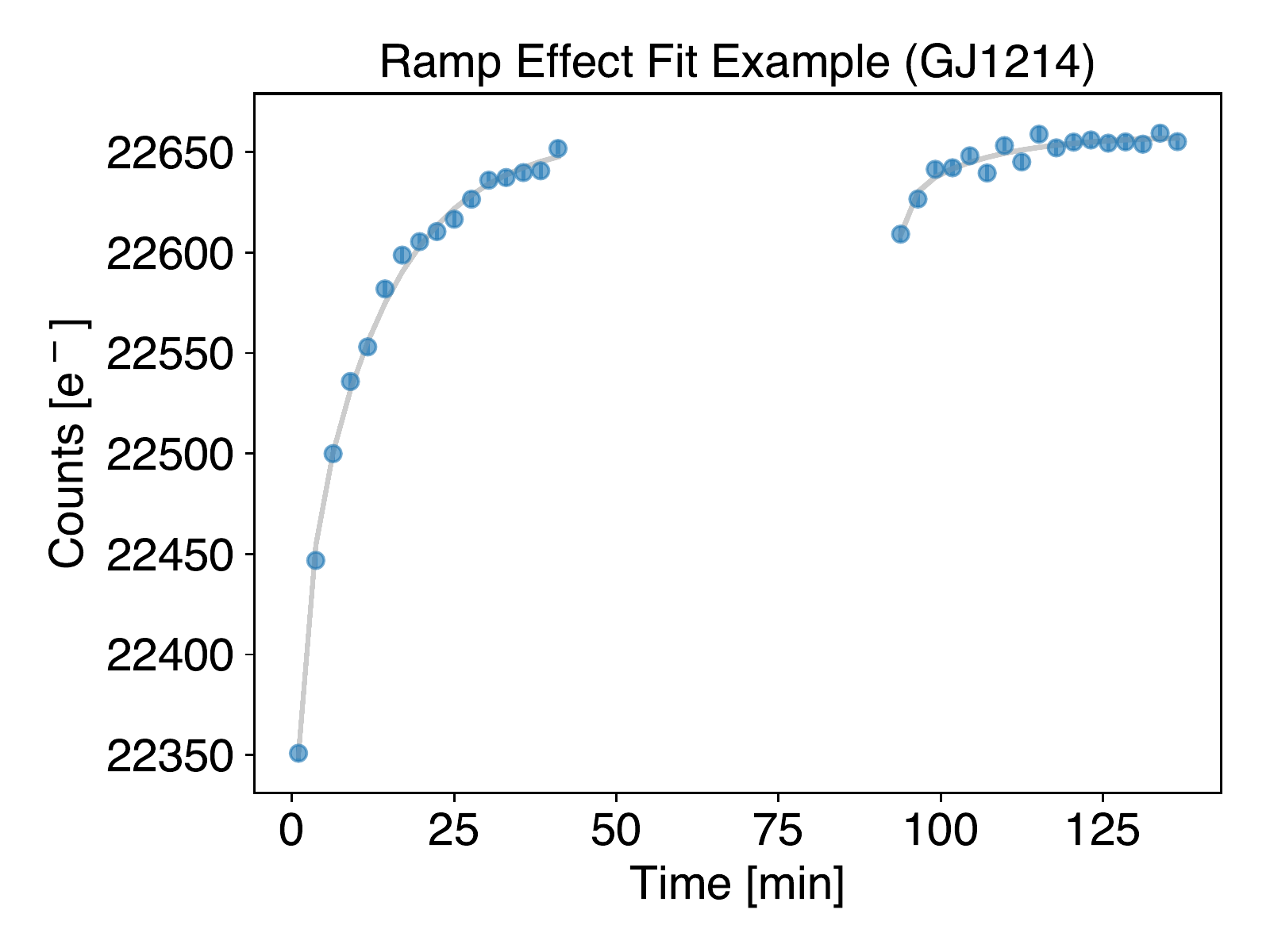}
  \caption{Example of the ramp effect light curve fitted to the
    RECTE model curve. Blue points are observational data
    that come from the average of one pixel column of the spatial
    scan. The gray lines are the RECTE model curve fits, each of which is one
    randomly chosen realization of the MCMC chain. The two
    orbits are well fit by to the ramp model light curve with
    maximum likelihood.}
  \label{fig:fit}
\end{figure}

The data reduction started with \texttt{ima} files produced by the
STScI \texttt{CalWFC3} pipeline that include all calibrated
non-destructive readouts, and followed the regular scanning mode data
reduction routine described in \citet{Deming2013} and
\citet{mccullough2012}. We marked the pixels that have data quality
flags of 4, 16, 32, and 256 (bad pixel, hot pixel, unstable response,
saturation) as invalid and excluded them from further analysis. Due to
the slight variability of the scanning rate and the jitter of the
telescope, the ramp effect shapes were buried under the noise of the
light curve of individual pixels. We averaged light curves of columns
of pixels along the scanning direction, so that the noise caused by
non-constant scanning rates was eliminated (Figure
\ref{fig:fit}). Therefore, we assumed the intrinsic light curves to be
a flat line for every column. We extracted 120 light curves for every
visit, and we had 1,800 light curves in total to fit the model
parameters. The signal-to-noise ratios for the extracted light curves are approximately 1000 per exposure.

\begin{figure*}[!th]
  \centering
  \includegraphics[width=\textwidth]{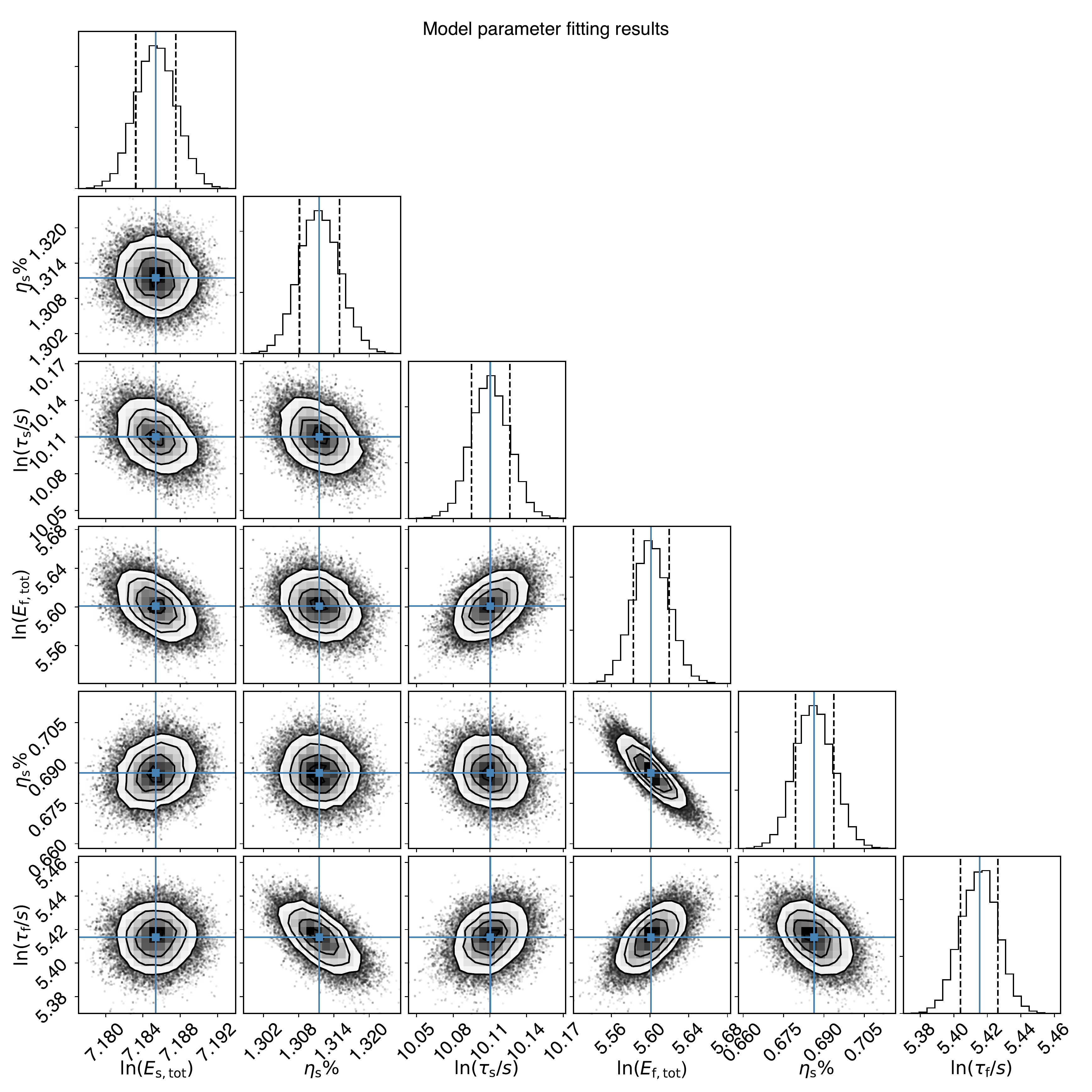}
  \caption{MCMC posterior distributions for $E_{\mathrm{s, tot}}$,
    $E_{\mathrm{f, tot}}$, $\eta_{\mathrm{s}}$, $\eta_{\mathrm{f}}$,
    $\tau_{\mathrm{s}}$, and $\tau_{\mathrm{f}}$. The blue lines mark
    the best-fit values while the black dashed lines show the
    1$\sigma$ uncertainty. The posterior distributions of
    $\eta_{\mathrm{s}}$  and $\eta_{\mathrm{f}}$ are normal, while the
    other four are log normal. There is no
    significant degeneracy among these three parameters except that
    between $\eta_{\mathrm{f}}$ and $E_{\mathrm{f, tot}}$.}
  \label{fig:mcmc}
\end{figure*}

We used a Markov Chain Monte Carlo (MCMC) procedure to find the
best-fit model parameters. The ramp effect profiles are determined by
11 parameters, including the flux count rate \flux, the initial number
of occupied traps $E_{0,\mathrm{s}}$ and $E_{0, \mathrm{f}}$, the
number of charge carriers captured between orbits
$\dtrap_{\mathrm{s}}$ and $\dtrap_{\mathrm{f}}$, and the 6 parameters
controlling the trapping processes ($E_{\mathrm{s, tot}}$,
$\eta_{\mathrm{s}}$, $\tau_{\mathrm{s}}$, $E_{\mathrm{f, tot}}$,
$\eta_{\mathrm{f}}$, $\tau_{\mathrm{f}}$). We initially allowed
  these 6 parameters to vary from column to column, but found that the
  values of the best-fit parameters from different columns agreed with
  each other within uncertainties. For the rest of the study, we
  therefore fixed the model parameters $E_{\mathrm{s, tot}}$,
  $\eta_{\mathrm{s}}$, $\tau_{\mathrm{s}}$, $E_{\mathrm{f, tot}}$,
  $\eta_{\mathrm{f}}$, and $\tau_{\mathrm{f}}$ to be the same for every
  input light curve. In this way, we focused on the average behavior
  of pixels and we were able to determine these properties more
  precisely. The likelihood function is expressed as
\begin{equation}
  \label{eq:likelihood}
  L =\prod_{i} \frac{1}{\sqrt{2\pi\sigma_{i}^{2}}} \exp\left(-\frac{[\mathrm{obs_{i}} -
      \mathrm{RECTE}(E, \eta, \tau, ...)]^{2}}{2\sigma_{i}^{2}}\right)
\end{equation}
The photometric uncertainty $\sigma$ is a combination of the photon
noise, read noise, \glenn{dark current}, and sky subtraction noise.
We assumed flat priors for \flux and $\eta$, and flat priors in $\log$
space for $N_{0}$ and $\tau$. For $E_{0}$ and \dtrap, we assumed an
exponential distribution as the prior distribution because, in most
cases, the initially occupied traps and extra added traps are close to
0. We sampled the posterior distributions and fitted 1,800 light
curves simultaneously so that the values of the trapping parameters
$E_{\mathrm{s, tot}}$, $E_{\mathrm{f, tot}}$, $\eta_{\mathrm{s}}$,
$\eta_{\mathrm{f}}$, $\tau_{\mathrm{s}}$, and $\tau_{\mathrm{f}}$ were
shared among all the light curves. In order to reduce the number of
free parameters, we first fitted each of the 1800 light curves to find
\flux, $E_{\mathrm{s},0}$, $E_{\mathrm{s},0}$, $\dtrap_{\mathrm{s}}$,
and $\dtrap_{\mathrm{f}}$. Second, we took the values of \flux,
$E_{\mathrm{s},0}$, $E_{\mathrm{s},0}$, $\dtrap_{\mathrm{s}}$, and
$\dtrap_{\mathrm{f}}$ calculated from the first step to fit for
$E_{\mathrm{s, tot}}$, $E_{\mathrm{f, tot}}$, $\eta_{\mathrm{s}}$,
$\eta_{\mathrm{f}}$, $\tau_{\mathrm{s}}$, and $\tau_{\mathrm{f}}$. We
iterated the two steps and found that \glenn{best-fit model parameters
  (Table \ref{tab:result})} quickly converged after just 5 iterations.

\section{Results}
\label{sec:results}

\begin{figure*}[t]
  \centering
  \includegraphics[width=\textwidth]{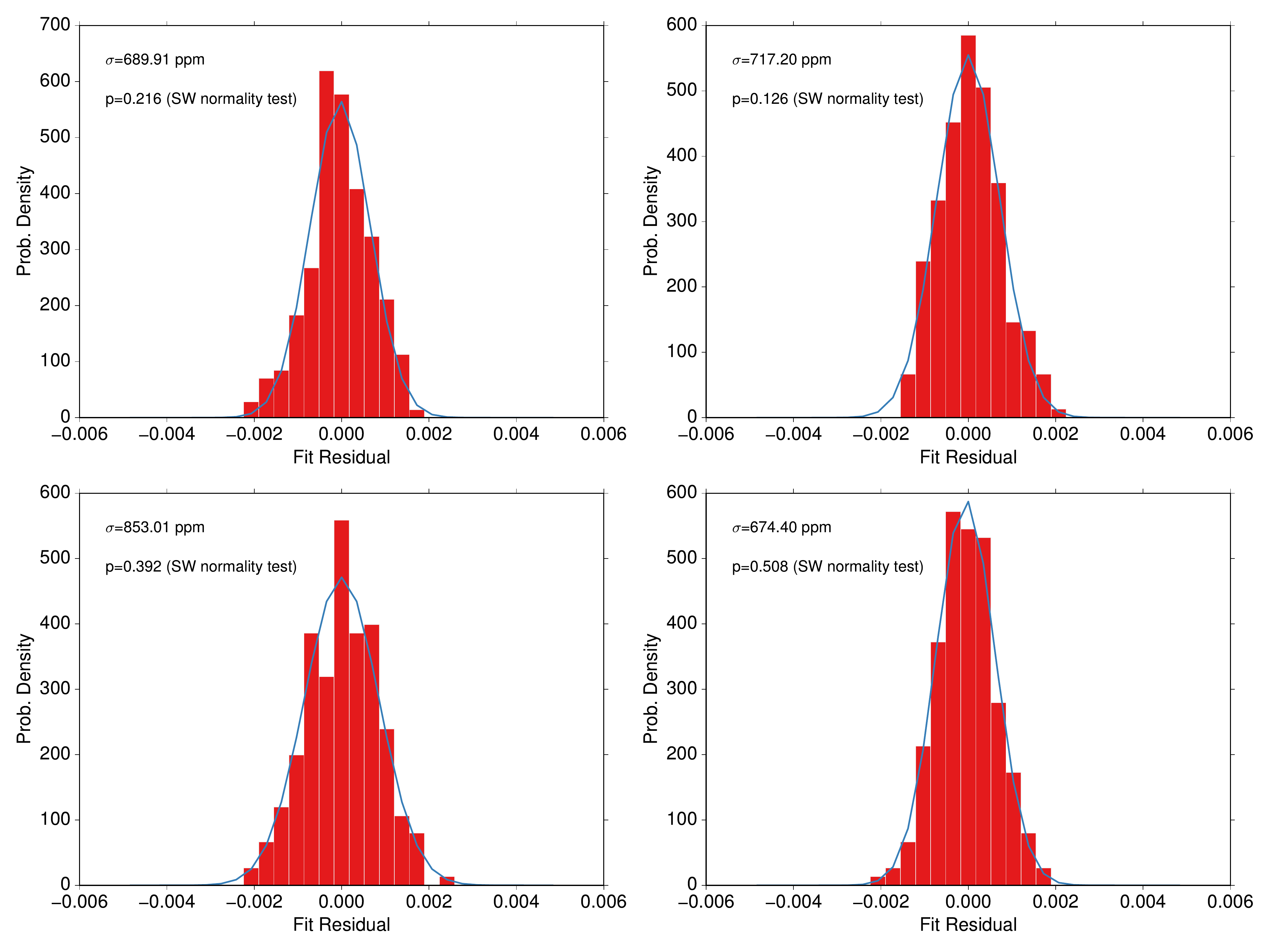}
  \caption{The distributions of light curve fitting residuals after RECTE model calibration. The similar distribution
    of the residuals for the 4 orbits alleviates the need of
    scheduling for extra orbits at the beginning of each visit for transit
    observations}
  \label{fig:residual}
\end{figure*}

\begin{figure*}[t]
  \centering
\includegraphics[width=\textwidth]{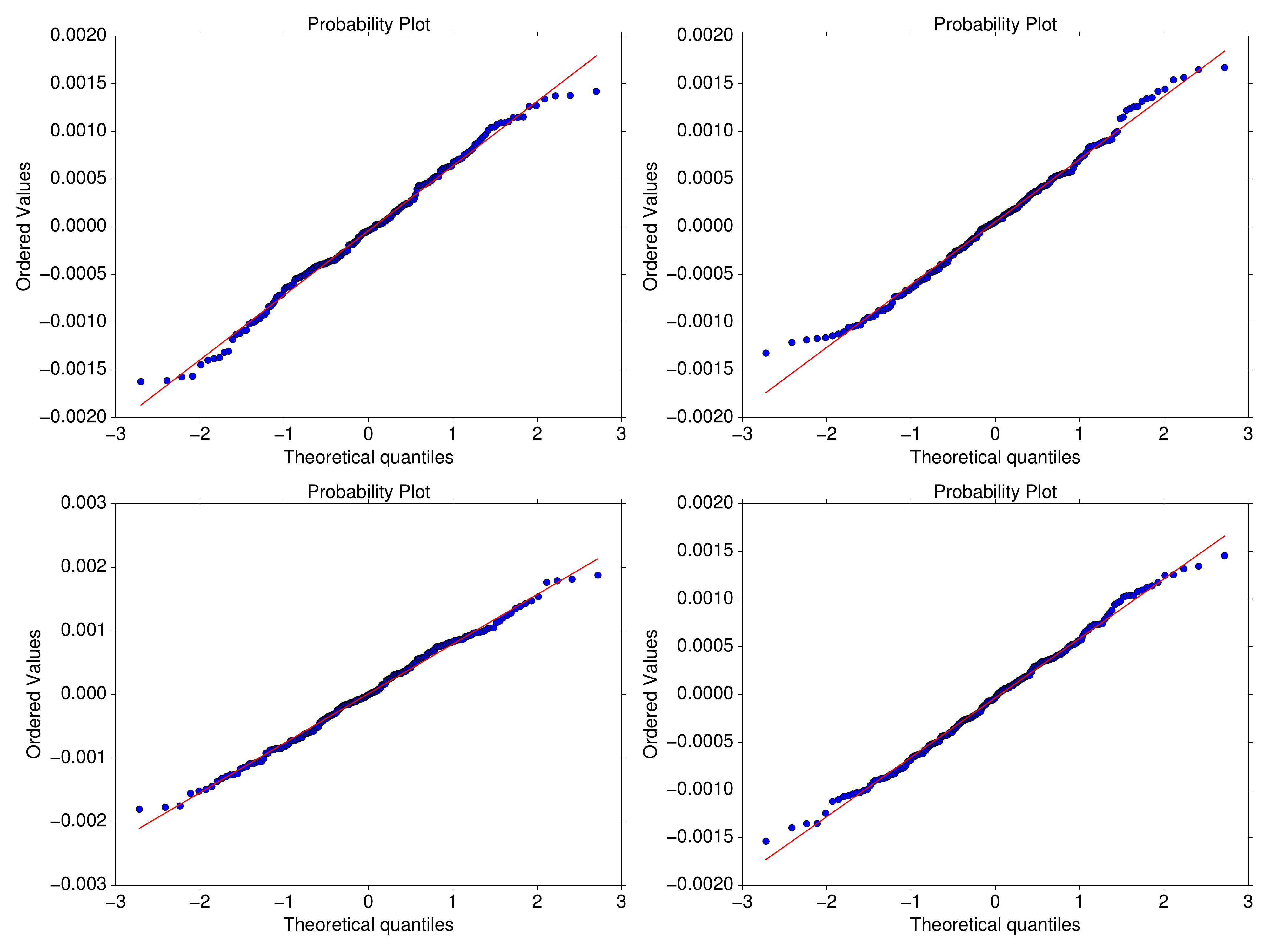}
  \caption{Quantile-quantile plot showing the fitting residuals have very little
    component of red noises in all 4 orbits. Quantiles of the residual
  are plotted against the quantiles of normal distributions. Very
  small deviations from linear relation demonstrate that red noises
  are mostly removed for all 4 orbits.}
  \label{fig:qq}
\end{figure*}

The result of the model fit is shown in Figure \ref{fig:mcmc}. The
model parameters are well constrained. The posterior distributions of
$E_{0}$ and $\tau$ are log-normal distributions, while the posterior
distribution for $\eta$ is Gaussian. We calculated the best-fit
values, as well as the 1-$\sigma$ uncertainties for the model
parameters, and list them in Table \ref{tab:result}. The fact that the
best-fit parameters are tightly constrained with small uncertainties
demonstrate that our model is able to make consistent correction for
every light curve in the entire dataset. We note that the best-fit
values for model parameters listed in Table \ref{tab:result} represent
average values for different pixels. Although small intra-pixel
variations in the trapping parameters would not be surprising, our
results demonstrate that a single set of trap is able to provide high
fidelity corrections for typical light curves.

\begin{deluxetable}{clcl}
\tablecaption{RECTE Model Fit Results \label{tab:result}}
\tablecolumns{4}
\tablenum{3}
\tablewidth{0pt}
\tablehead{
\colhead{Parameter} &
\colhead{Value}&
\colhead{Parameter} &
\colhead{Value}
}
\startdata
$E_{\mathrm{s,tot}}$&$1320.0\pm2.8$&$E_{ \mathrm{f, tot}}$&$270.6^{+5.2}_{-4.9}$\\
$\eta_{\mathrm{s}}$&$1.311\pm0.0034$\%&$\eta_{\mathrm{f}}$&$0.6863\pm0.0070$\%\\
$\tau_{\mathrm{s}}$&$2.45^{+0.040}_{-0.037}\times 10^{4}\mathrm{s}$&$\tau_{\mathrm{f}}$&$224.8^{+2.6}_{-2.4}\mathrm{s}$\\
\enddata
\end{deluxetable}

The additional 4 parameters, $E_{0, \mathrm{s}}$, $E_{0, \mathrm{f}}$,
$\dtrap_{\mathrm{s}}$, and $\dtrap_{\mathrm{f}}$ vary on a visit to
visit basis. $E_{0, \mathrm{s}}$, $E_{0, \mathrm{f}}$ varies from 0 to
20\% of the total numbers of traps, representing the uncertainty of
the initial state of the detector. The values for
$\dtrap_{\mathrm{s}}$, and $\dtrap_{\mathrm{f}}$ are normally close to
0, but help to provide finer fit early in the orbit.

We tested the validity of the model using the 4-orbit-long transits of
the 12 visit observations. As an example, we plotted the distributions
of the model fit residuals in 4 orbits of a one pixel column in Figure
\ref{fig:residual}. The residual distributions of the 4 orbits are all
close to Gaussian distributions, and have standard deviations of less
than 1.1 times of photon noise. We did Shapiro-Wilk normality tests
for all 4 orbits observations of every one-pixel column. The residuals
for 4 orbits of every pixel column agree with normal distributions
with a $\alpha=0.1$ fidelity. We further tested the normality with
quantile-quantile plots as shown in Figure \ref{fig:qq}, where we
plotted the observed quantile against the quantile following Gaussian
distributions. The residuals agree with straight lines for all 4
orbits, which further verifies that our model successfully corrected
 the ramp effect for all 4 orbits.

 We further tested our model with datasets listed in Table
 \ref{tab:params} to demonstrate that our model works for observations
 done using different telescope and instrument configurations with
 different levels of incoming flux.

\subsection{Applications}\label{sec:app}

We found that our model works efficiently for a wide variety of \WFC
time-resolved observations with different instrument modes and
targets.
\subsubsection{Scanning and staring mode observations}
For scanning mode grism observations where the scanning direction is
perpendicular to the dispersion direction, a high signal-to-noise light
curve for one wavelength element can be obtained by summing up
the pixel counts along the scanning direction. Random noise introduced
by non-constant scanning rates and telescope jitter are mostly
eliminated by this step.  We directly applied RECTE to the
light curve extracted from each pixel column to remove the ramp effect.

Compared to scanning mode observations, staring mode observations have
two major differences for the application of RECTE. First,
within a region of interest, pixels are illuminated at vastly
different levels. Because the ramp effect profile is related to the
pixel count levels, the total ramp effect within the region of
interest cannot be calculated as the average of pixels.  Second,
light curves for individual pixels may be affected by telescope
jitter. With \HST's excellent pointing stability (typically less than
0.1 pixel drift per orbit), telescope jitter has little effect on the
overall ramp profiles. We compared the ramp effect for light curves of
simulated G141 staring mode observation images with no pointing
jitter/shift and with pointing shifts of 0.1 pixel per orbit, and
found that the relative difference is less than 0.01\%. Therefore, for
staring mode and imaging observations, the ramp effect can be corrected by
applying RECTE without taking pointing errors into account.

\subsubsection{Transiting planet example: GJ1214}

\begin{figure}[th]
  \centering
  \includegraphics[width=\columnwidth]{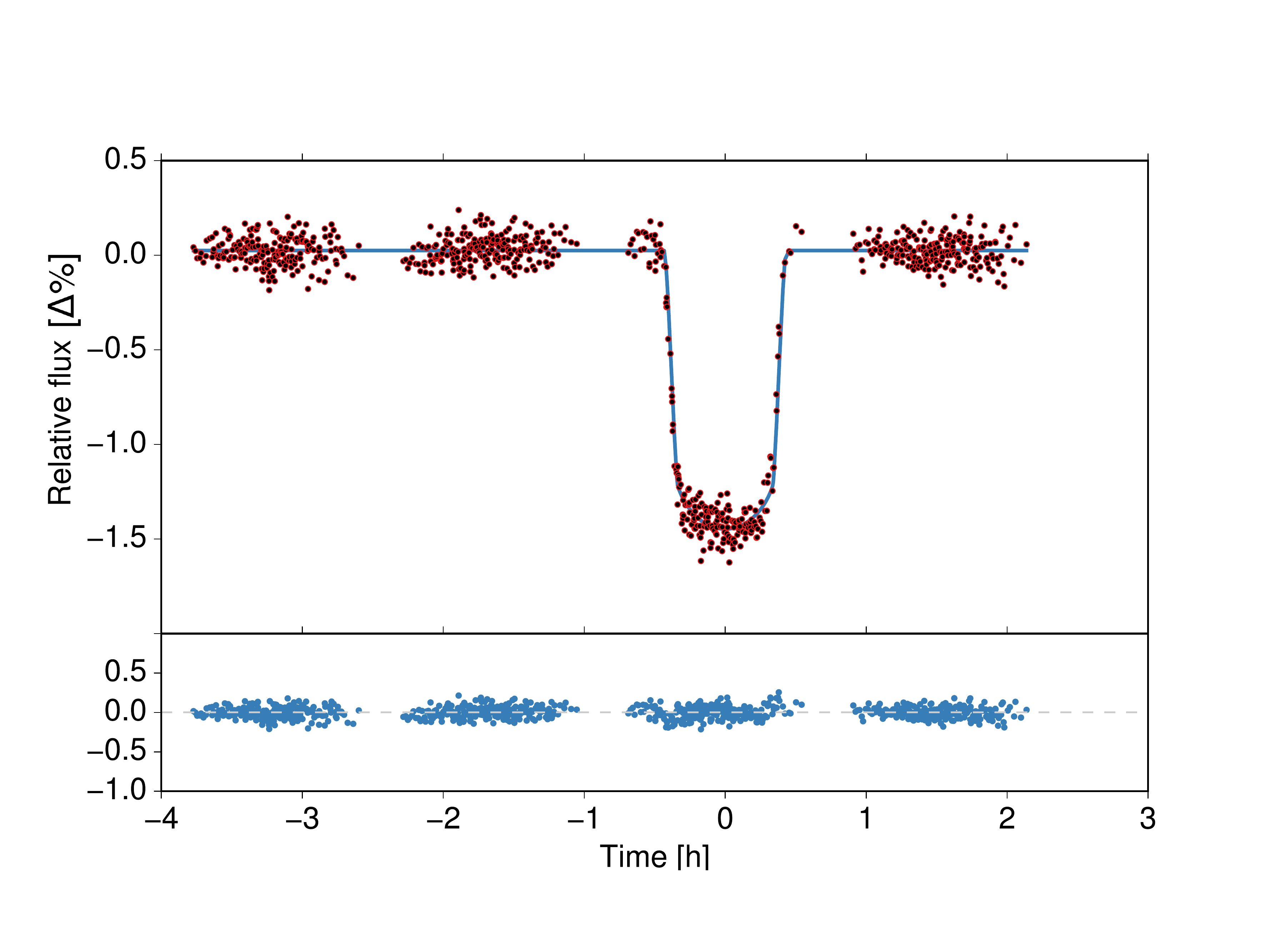}
  \caption{4 orbits of GJ1214 transit observations are well fitted by RECTE+transit profile model.}
  \label{fig:transit}
\end{figure}

\begin{figure*}[th]
  \centering
  \plottwo{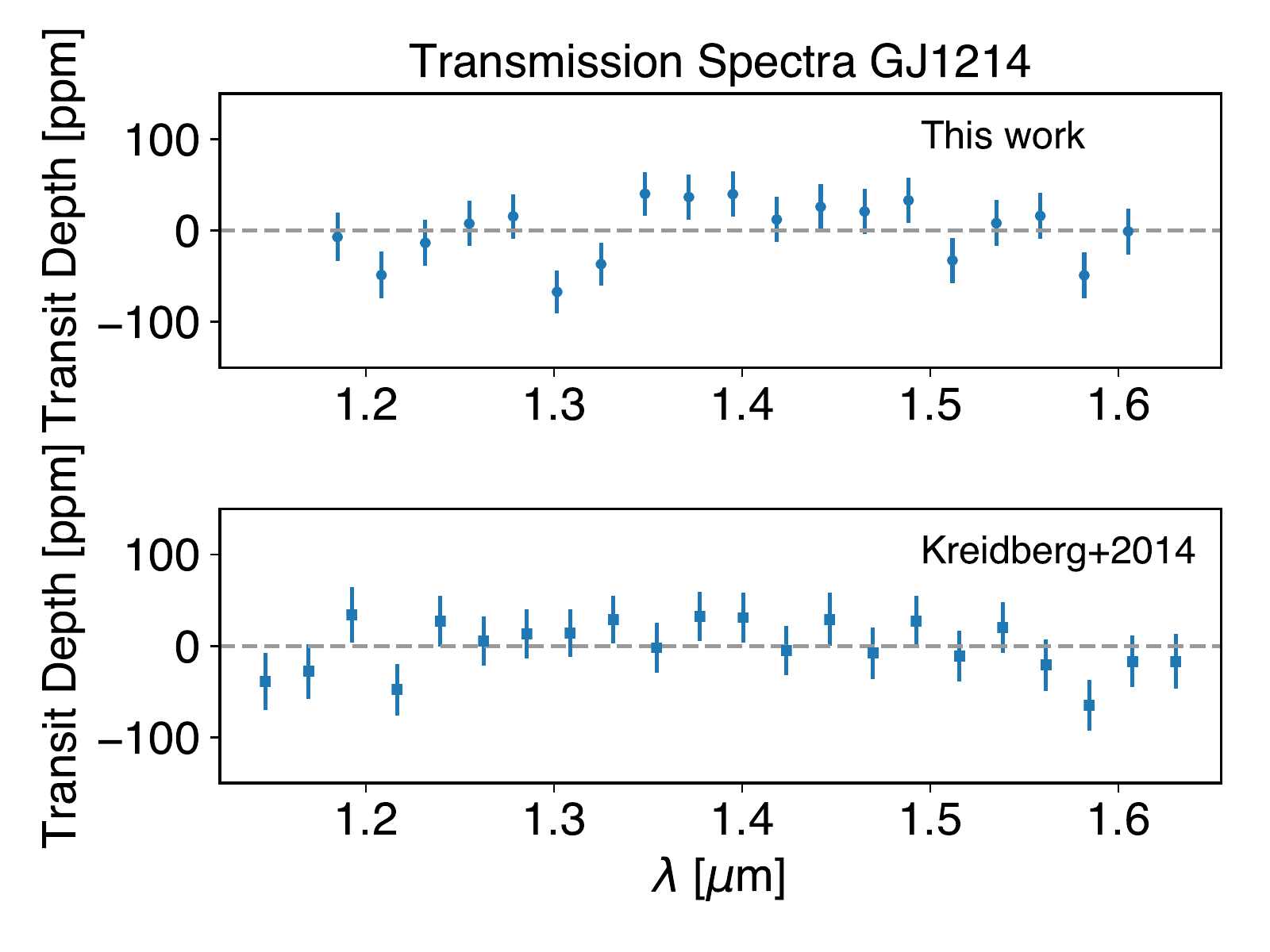}{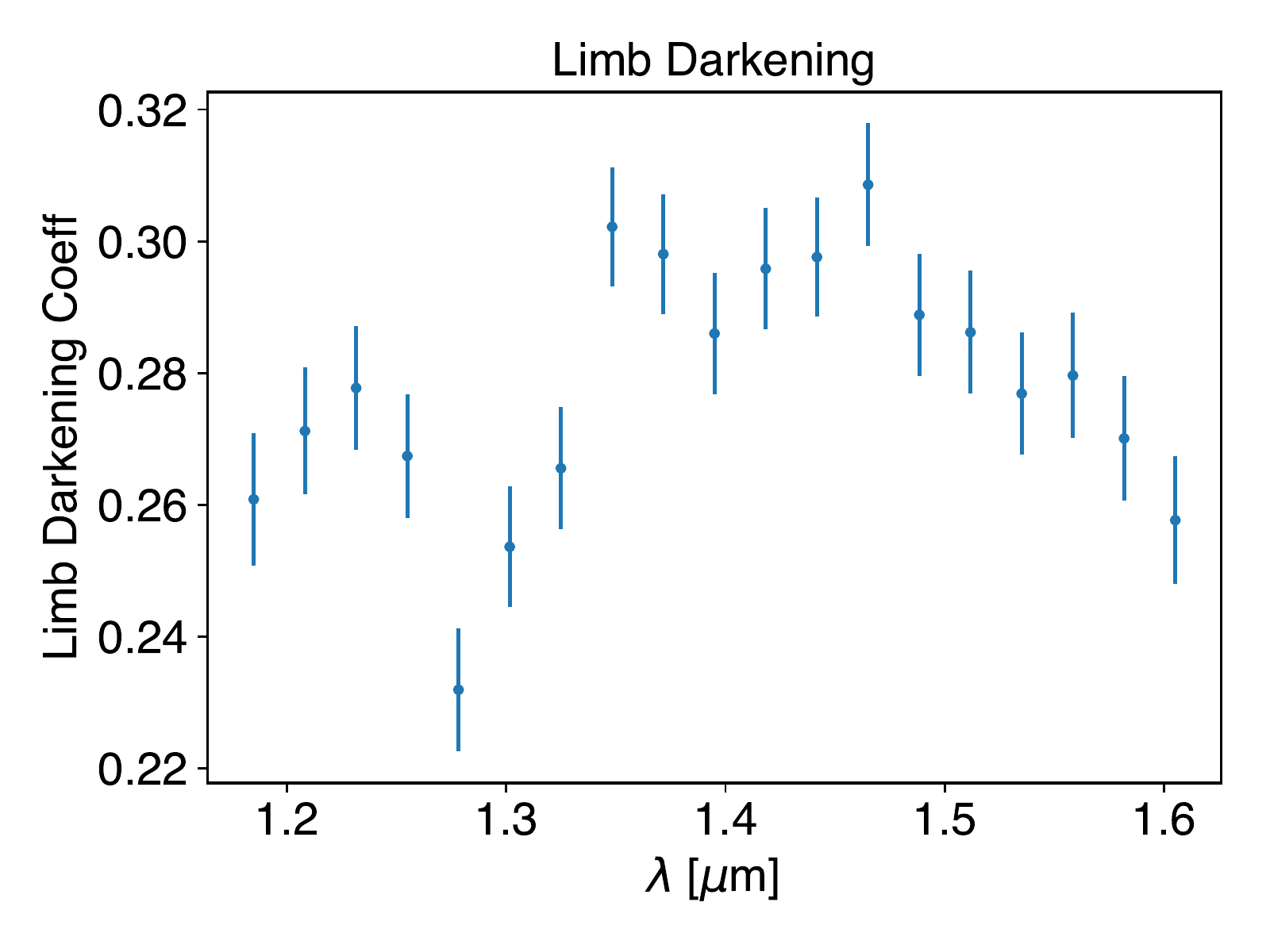}
  \caption{Transmission spectra (left) and limb darkening
    coeffecient (right) measured from this work comparion to that of
    \citet{Kreidberg2014}. The measurements agree well within
    uncertainties.}
  \label{fig:spec}
\end{figure*}

\begin{figure}[h]
  \centering
  \plotone{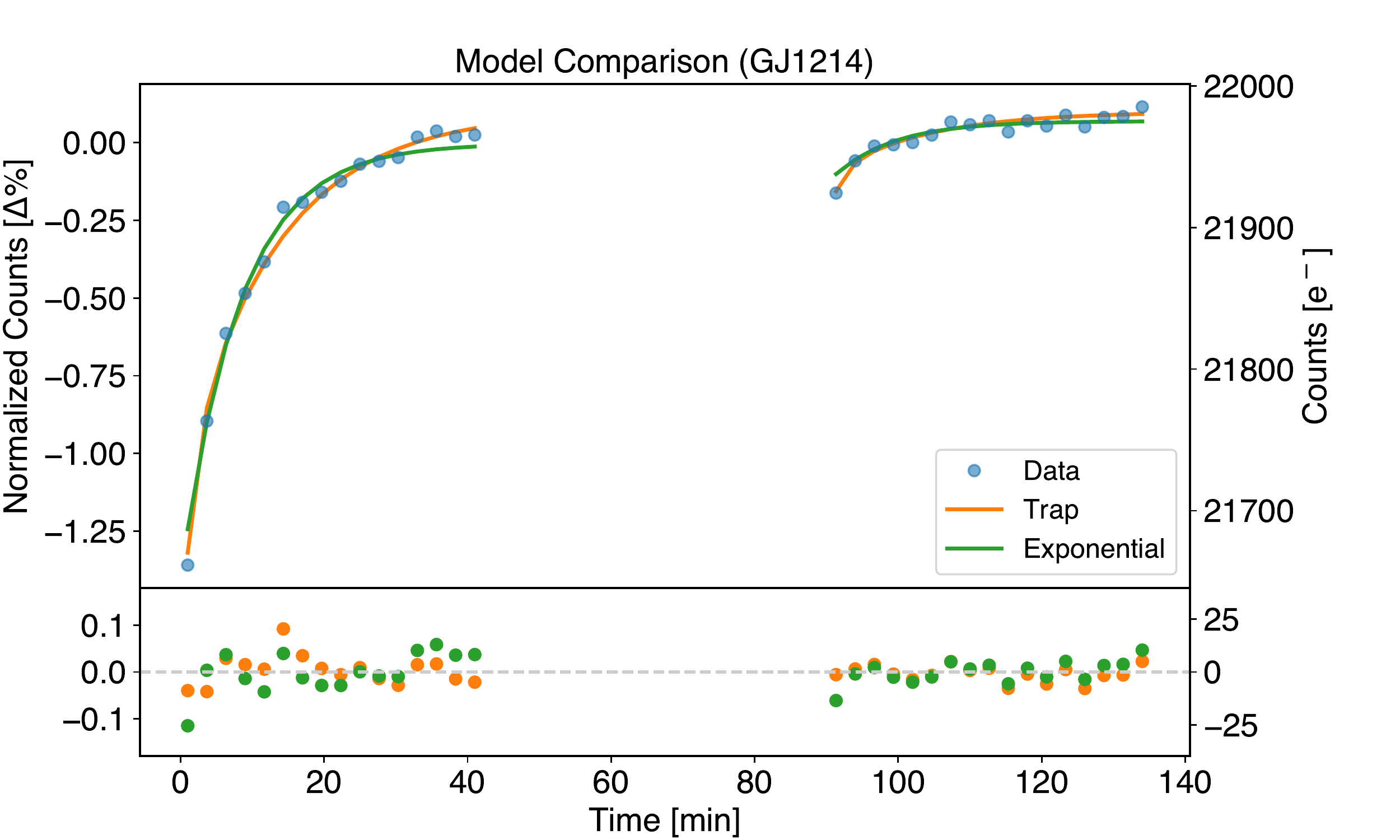}
  \caption{Comparison of systematics corrections and fitting
      residuals for one light curve from GJ1214 dataset. The orange
      curves and dots are ramp profiles calculated using the proposed
      model and its fitting residuals, and the green curves and dots
      are those for the best exponential fit using the white light
      curves. Because the change of ramp profiles across different
      wavelengths, the best-fit exponential profile failed to fit the
      beginning of both orbits.}
  \label{fig:comp}
\end{figure}

While testing the validity of the model, we also measured the
transmission spectrum (Figure \ref{fig:transit}) of GJ1214b. We
reached the same level of correction precision as the
divide-out-of-transit method used by \citet{Kreidberg2014}, and
obtained a very similar transmission spectrum (See Figure
\ref{fig:spec}). We note that for a few wavelength channels, the
normal divide-out-of-transit method shows limitations. In Figure
\ref{fig:comp}, we compared the ramp profiles provided by our model
and the best-fit exponential profile in broadband light curve. The
exponential fit fails to reproduce the pattern in the light curves at
the beginning of the first and second orbits, while our RECTE model
provides matches those well. In conclusion, we find that the RECTE
model matches well the accuracy of the best empirical correction yet
achieved HST (taking in observations designed with the empirical
correction in mind), but our observations do not rely on extra orbits
to reduce the ramp effect.

GJ1214b \citep{Charbonneau2009} is a 6.5 $M_{\oplus}$ transiting
planet. The large transit depth and brightness of the host star make
it one of the best-suited sub-neptunes for transmission spectroscopic
observations. This planet has been extensively observed both with
space-based and ground-based facilities in multiple wavelengths from
optical to mid-infrared
\citep[e.g.,][]{Bean2011,Bean2010,Fraine2013}. The data that we used
to constrain and validate the RECTE model were originally
published by \citet{Kreidberg2014}, where they found a flat near
infrared transmission spectrum for GJ1214b between 1.15 and 1.65
$\mu$m. The featureless transmission spectrum requires high-altitude
clouds/haze in the atmosphere of GJ1214b to suppress the otherwise
strong water absorption features. For their data reduction,
\citet{Kreidberg2014} discarded the first orbit of every visit due to
the different shapes of the ramp effect profiles. They used both
empirical exponential functions and white light curves to correct the
ramp effect, and obtained a consistent transmission spectrum and
limb-darkening profile (Figure \ref{fig:spec}). The consistent results
of this work and of \citet{Kreidberg2014} confirms that GJ1214b has a
featureless near-infrared transmission spectrum. With RECTE,
future observations will not need to discard the first orbit
observations, which will significantly improve the efficiency of \WFC
observations by about 25\%.

\subsubsection{Scanning mode Observations: attempting to mitigate the
  ramp with tungsten lamp pre-conditioning}
\begin{figure*}[th]
  \centering
  \includegraphics[width=0.33\textwidth]{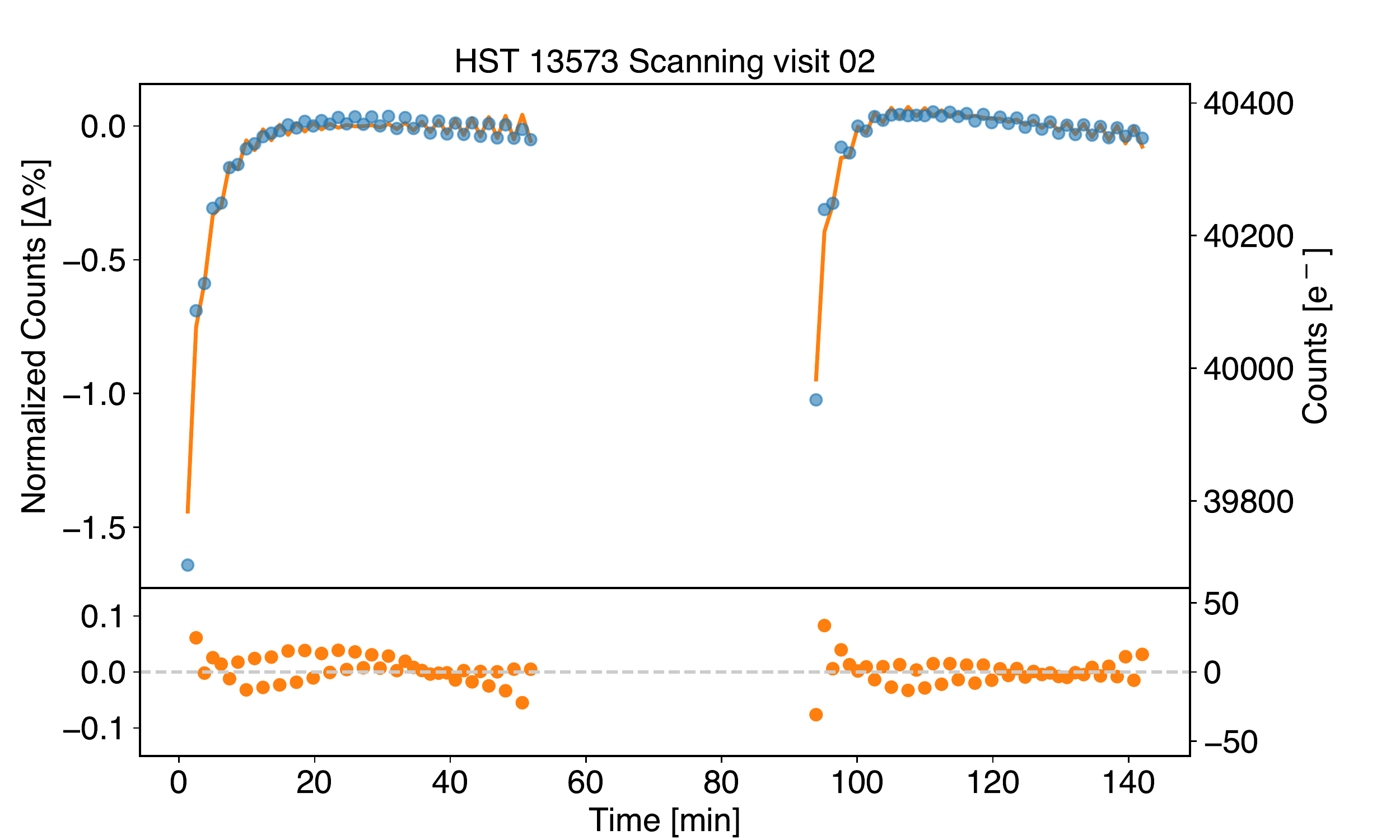}
  \includegraphics[width=0.33\textwidth]{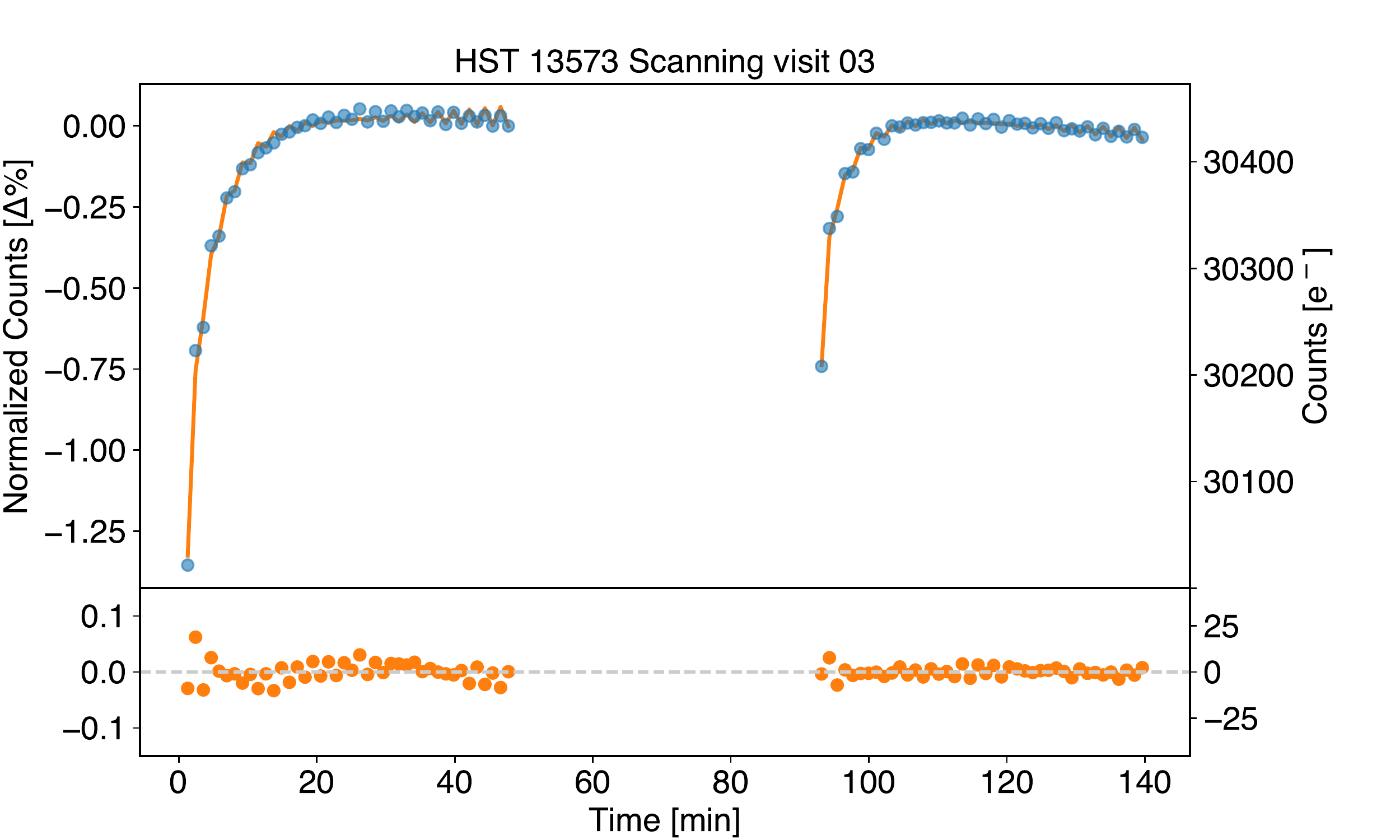}
  \includegraphics[width=0.33\textwidth]{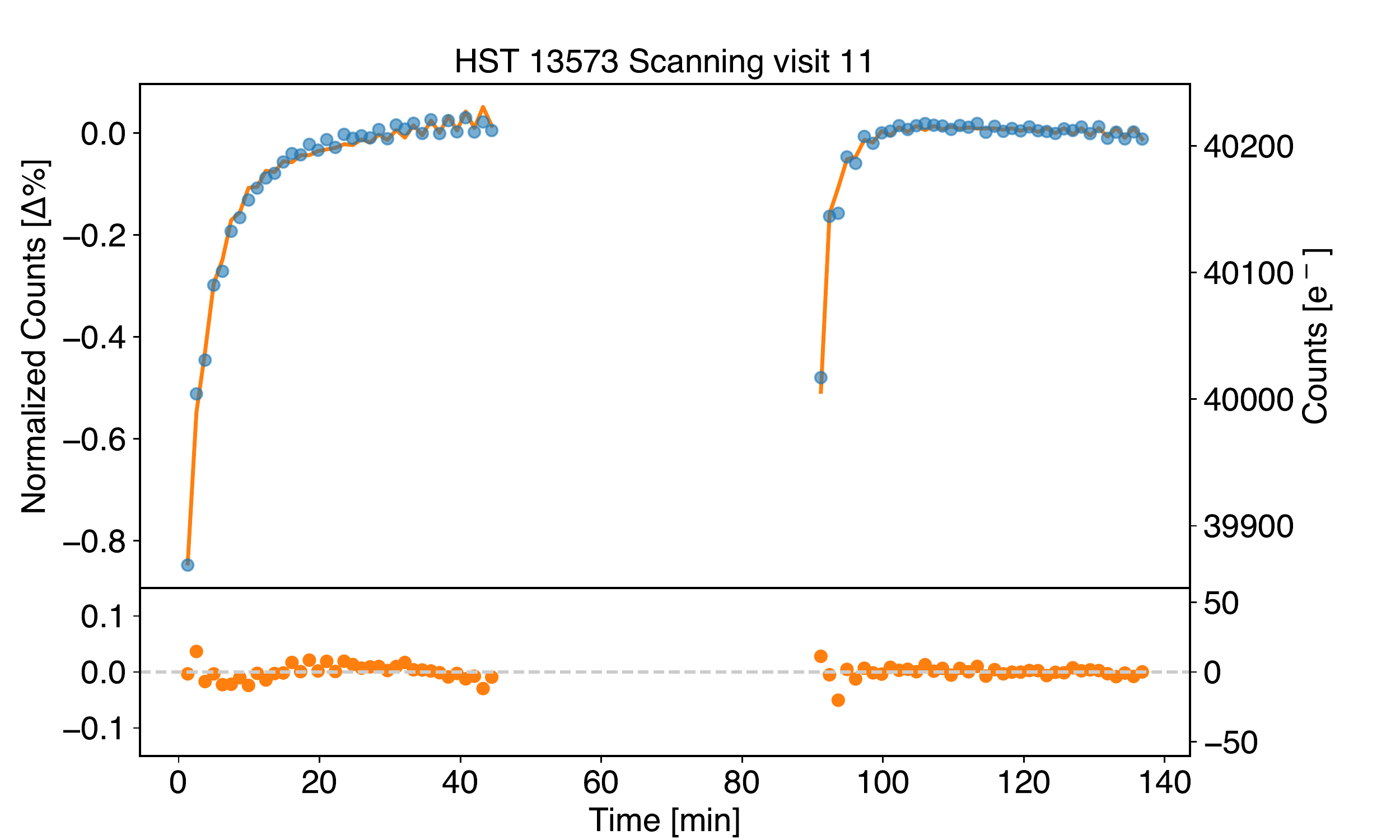}
  \caption{Applying RECTE model to light curves taken in the
    experiment to mitigate the ramp effect with
    pre-conditioning. Pre-conditioning using the tungsten lamp
    demonstrated very limited ability to fill the charge carrier traps
    (left and middle panel), while taking exposures with the source
    shows better result (right).}
  \label{fig:mitigate}
\end{figure*}
\citet{Long2014} attempted to mitigate the \WFC ramp effect by
pre-conditioning the detector. They took exposures with a tungsten
lamp within the instrument to fill up the traps. They turned on
the lamp for $\sim30$ seconds, and kept the count level for
$\sim2,600$ seconds to attempt to saturate the proposed charge carrier
traps. The amplitudes of the ramp effect was somewhat reduced in their
experiment, but they could not fully remove the ramp effect via
pre-conditioning.

We applied our model to the light curves acquired in their
experiment. Our model can well reproduce the ramp profiles for the
detector pre-conditioned with the tungsten lamp. The results are shown in
Figure \ref{fig:mitigate}. With our model, we found that
pre-conditioning using the tungsten lamp has a very limited effect on
filling up the charge carrier traps, while using the source itself to
illuminate the detector did have more significant effect on filling
the traps.  Comparing the number of filled traps after pre-conditioning
with tungsten lamps with that after taking an orbit-long exposure
series, we found that keeping the detector pixels at high electron
levels is inefficient to fill the charge carrier traps. In the
experiment of \citet{Long2014}, there is one visit for which the
tungsten lamp was inadvertently left on during the whole pre-condition
period. The filling of the charge carrier traps appeared to be more
efficient for that visit because the ramp profiles for the first and second
orbits have very similar amplitudes and shapes (see Figure 3 of
\citet{Long2014}). However, the light curve of that visit demonstrated
additional systematics that could be related to long-lasting
illumination by the lamp.

\subsubsection{Rotational phase mapping: 2M0915}

\begin{figure}[th]
  \centering
  \includegraphics[width=\columnwidth]{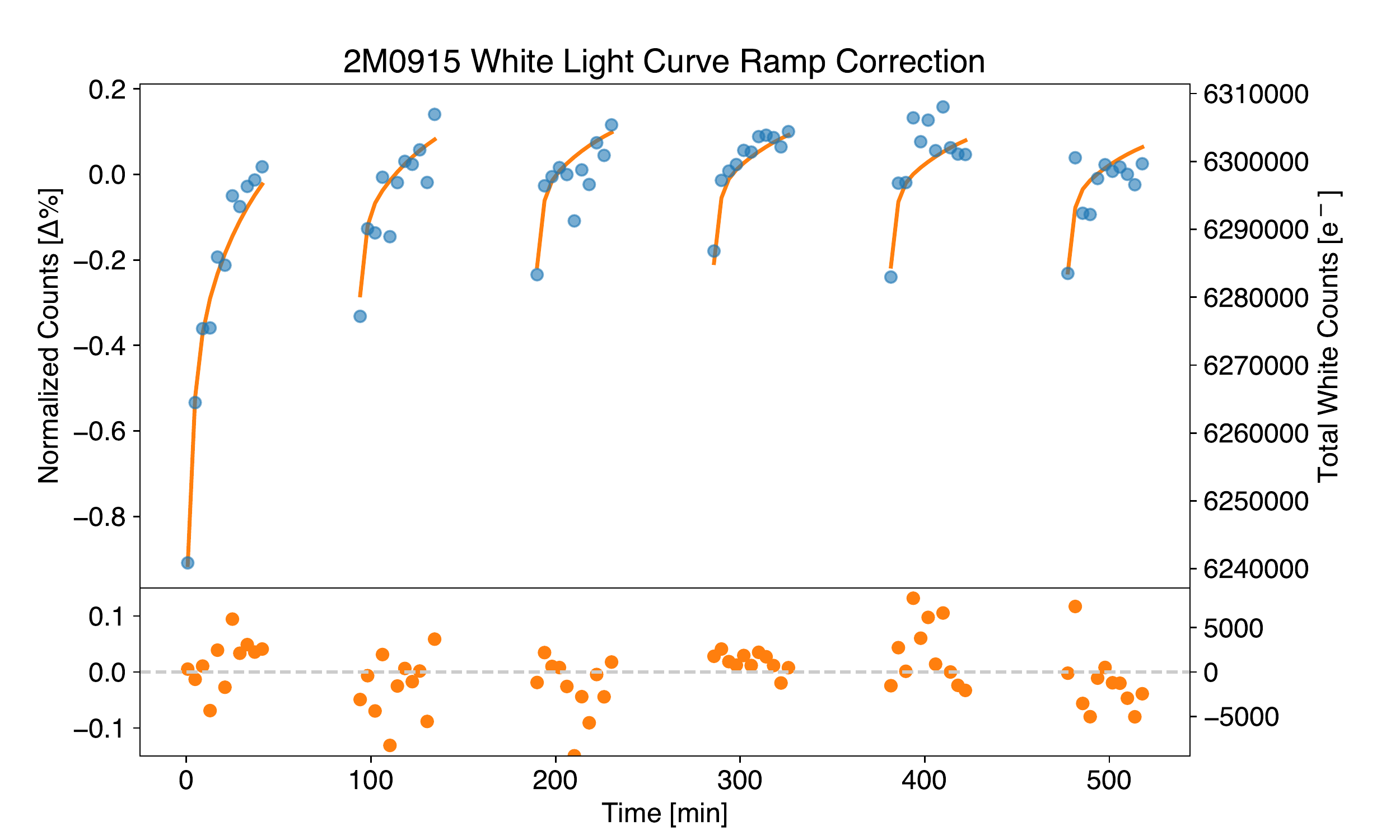}
  \caption{6 orbits observations of binary brown dwarf binary 2M0915
    corrected by RECTE}
  \label{fig:2M0915}
\end{figure}

We used our model to correct \WFC time-resolved observations of the
binary brown dwarf system 2M0915 (\HST program 12314, P.I. Apai) as an
example to demonstrate the model's application to staring mode
observations. 2M0915 system has two L7 type brown dwarfs with an
angular separation of $0.7''$ \citep{Reid2006}. The observations
included 6 orbits of G141 exposure series (SPARS25, NSAMP=12 exposure
time 223.73s). Each orbit had 11 exposures. For the following
demonstration, we focus on the combined light curve of the binary and
do not attempt to separate the two grism spectrum components. The raw
light curve for the entire G141 grism wavelength span (white light
curve) shows a prominent ramp effect profile in every orbit with an
average amplitude of 0.5 \% (Figure \ref{fig:2M0915}), more than 10
times of the photometric uncertainty.

We corrected the ramp effect with the RECTE model. The
corrected light curve agrees fairly well with a flat light curve with
a reduced $\chi^{2}$ of 1.76 (63 degrees of freedom). The fitting
residual of straight line with RECTE model applied has a
standard deviation of 0.055\% (1.1 $\times$ the photon noise).

Rotational phase mapping from time-resolved spectroscopy provides the
most direct observational constraints on the condensate clouds of
ultra-cool atmospheres \citep[e.g.,][]{Apai2013,Buenzli2015}. The
photometric and spectral modulations are introduced by heterogeneous
clouds whose projected surface area is modulated by the rotation of
the objects. The survey conducted by \citet{Buenzli2014} concluded
that at least one-third of brown dwarfs with spectral types from mid-L
to mid-T have observable rotational modulations in the near-IR band
from 1.1 to 1.7 $\mu$m. Using Spitzer observations,
\citet{Metchev2015} also detected rotational modulations for 30\% to
40\% L3-L9.5 brown dwarfs in \textit{Spitzer} 3.6 and 4.5 $\mu$m
band. They both claimed that almost all brown dwarfs have
heterogeneous cloud coverages, after taking into account the random
distributions of the inclinations of rotation axis and limited time
coverage and sensitivity of their observations.

The corrected light curve of the brown dwarf binary 2M0915 agrees well
with a flat light curve. Based on the corrected light curve, the
possibility of rotational modulation with peak-to-peak amplitude
larger than \glenn{0.15\% with a period shorter than 2.5 hours can be excluded
  above 3-$\sigma$ level.}
The flat combined light curve of the
binary brown dwarf can be interpreted as either that the two binary
components having very homogeneous atmospheres, or the spin axes of both
of the two brown dwarfs having very close to pole-on inclinations.  The
first interpretation is unlikely, given the high occurrence rate of
heterogeneous atmospheres. Therefore, both of the two brown dwarfs
should have nearly pole-on spin axes. Constraints of the orbital
motion of the binary system will help study the orbit-spin alignment
and the dynamical evolution in this system.

\section{Discussion}
\label{sec:discussion}

\subsection{Mitigating the ramp effect}\label{sec:suggestion}
\citet{Swain2013} proposed several methods for time-resolved
observations using \WFC to reduce the ramp effect, including using
small sub-array mode and keeping low count levels
($<40,000\, \mathrm{e}^{-}$). Our model quantitatively supports some suggestions.
\glenn{The data storage memory of \WFC only can only save 32 full
  ($1014\times1014$) near IR channel frames, so high-cadence
  time-resolved observations with array sizes of $512\times512$ or above
  often require multiple buffer downloads within the visibility period.
  The buffer download time of \WFC is $8-9$ min, which is more than 5
  times the lifetime of the fast traps.}
Therefore, the fast traps will release a considerable number of
charges and be available again during the buffer download time,
leading to a more significant ramp effect. As a
comparison, the buffer download times for $256\times256$ and
$128\times 128$ sub-arrays are similar to or even shorter than the fast
trap lifetime. Observations taken in these two small sub-array
modes normally do not require buffer downloads in the middle of the
orbits. Therefore, there are no significant breaks in the exposure
series during which trapped charge carriers can be released but not
added during orbits when the smaller sub-arrays are used. \glenn{As
  for the recommendation for keeping low count levels, we see no
  evidence for the need to calibrate the ramp effect differently for
  different exposure levels (i.e., the same trap parameters provide
  excellent fits).} Observations with relatively low count levels
(e.g., 2M0915) and high count levels (e.g., HD 209458) have similar
ramp effect amplitudes with different apparent ramp effect profile,
and can be corrected by the same model.

When applying RECTE to correct the ramp effect, the only
degrees of freedom of the ramp profiles are the initial numbers of the
charge carrier traps, and the extra numbers of charge carriers trapped
during the gaps between the orbits. The degrees of freedom of the
model can be further decreased to acquire more accurate ramp effect
correction, if the initial states and telescope pointing information
between the orbits can be recorded. Therefore, for future
  observations, we suggest that detector images should be recorded
  before the telescope is reset prior to the science observations to
  allow measuring the illumination levels the detector was exposed to
  during the inter-orbit gaps. Alternatively, or in addition, the
detector's inter-orbit exposure should be minimized during
high-precision time-resolved observations: Because the WFC3/IR detector has
no mechanical shutter, we recommend that the filter wheel is set to a
``blank'' (opaque blocker) position, also used to prevent the detector
from viewing Earth during occultations.

\subsection{The relationship between the number of traps and fluence levels}

The theories of charge carrier trapping \citep[e.g.,][]{Smith2008a}
suggest that the total number of charge carrier traps can increase if
the illumination fluence levels are very high. The width of the
undepleted regions (see Figure 1 of \citealt{Smith2008a}) increases
with fluence level, which potentially enlarges the number of available
traps. Indeed, \citet{Long2012, Long2015a} showed that the persistence
of the \WFC IR detector surged when fluence reached near saturation,
which could suggest significant trap density increase at
saturation. However, for the scientific cases that this study focuses
on, in which the maximum fluence levels are almost always kept well
below the saturation, we see no evidence for this effect, and a
constant level of available traps reproduces the observations very
well.

\begin{figure}[ht]
   \centering
   \includegraphics[width=\columnwidth]{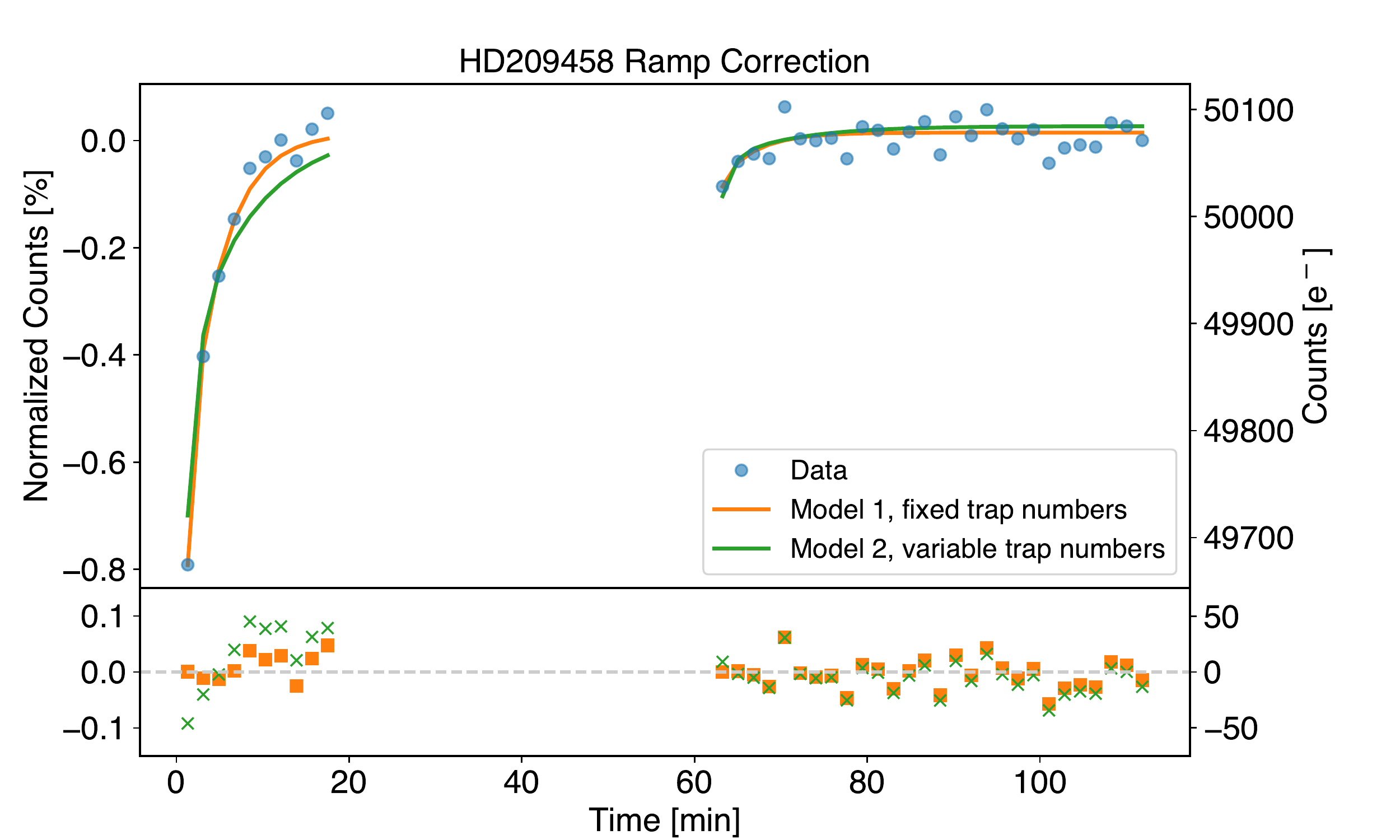}
   \caption{The light curve of first two orbits of scanning
       mode observations of transiting exoplanet HD209458b compared
       with two model curves. Despite the different incoming flux
       compared to GJ1214b the ramp effect trend can be corrected with
       high accuracy with model parameters exactly the same as those
       fitted using the scanning mode observations of GJ1214b. In
       contrast, the model (green) that assumes that the number of
       traps is directly proportional to the fluence level predicts a
       light curve that is inconsistent with the observations.}
   \label{fig:HD209458}
 \end{figure}

Figure \ref{fig:obsRamp} demonstrates a qualitative trend by which the
  the first orbit light curve ascends faster with increasing count
  level and flux. For example, from Figure \ref{fig:obsRamp} panel A to D, as
  the flux gets lower, the slope of the first 5 points also decreases
  (0.10[\%/min], 0.088[\%/min], 0.046[\%/min], and 0.034[\%/min],
  respectively.) If we assume that the number of traps are
  proportional to the fluence level, given a fixed exposure time, the
  exponential index in equation \ref{eq:4} and the ramp profiles would
  be independent of illumination flux, which contradicts to the
  observed trend. To illustrate this point, we compare the first two
  orbits of the scanning mode observation of HD 209458 (Figure 1A, a
  transiting hot-Jupiter host; \citealt{Deming2013}) to two model light
  curves as shown in Figure 12. Note that the average fluence level
  for HD 209458 (Fig 1A) is more than twice that of GJ1214 (Fig
  1B). In the first model, we use the same model parameters as the
  best-fit parameters for GJ1214. In the second model, we
  proportionally increased the number of traps for both populations
  based on the different fluence levels of two observations. As shown
  in Figure \ref{fig:HD209458}, the second model cannot fit the steep
  ramp in the first orbit as well as the first model does, resulting
  in fitting residuals twice as large as those from the first
  model. Therefore, we conclude that for observations with fluence
  levels well below saturation, there is no benefit from considering
  varying trap numbers for ramp effect calibrations. 

 \subsection{Ramp effect with image persistence in \WFC IR}

 The idea of charge carrier trapping originated from studies of
   image persistence. A complete model of charge carrier trapping
   should naturally explain both the ramp effect and image
   persistence. From an empirical perspective, a series of studies by
   \cite{Long2012,Long2015a,Long2015b} provided the most accurate
   model to predict image persistence for \WFC IR detectors yet. This
   model is publicly available as the ``persistence pipeline.'' The
   key characteristics of the model of Long et al. include a
   significant rise of persistence for fluence near saturation levels,
   persistence decaying as a power law, and persistence being related
   to the exposure times. In contrast, the model presented here does
   predict that persistence varies with exposure time, but saturation
   is not relevant in our model, and the persistence decays
   exponentially after the end of the exposures. For fluence below
   50000 $\mathrm{e^-}$ (2/3 of saturation level), our model predicts
   persistence of less than 0.05 $\mathrm{e^-}$ after 500 seconds of the
   exposure, which is at the similar order to the observational
   measurements of \cite{Long2012, Long2015a}.

 The differences between the two models reflect their different
   goals: our model was developed to correct the ramp effect in
   high-cadence, short exposure time observations, while the model by
   Long et al. corrects for persistence in exposures that follow long
   integrations and fluence levels close to saturation.
   For the observations that our model focuses on, the trapping of the
   charge carriers normally reduces the observed flux by
   $\sim0.2-1.5\%$, while the increase of the observed flux from
   persistence (released charge carriers) is two orders of magnitude
   lower. The persistence profile in \cite{Long2012, Long2015a} can be
   qualitatively explained with a third population of traps that has
   a broad range of trapping lifetimes that are only activated when
   fluence is above a certain threshold. Combining the approaches by
   Long et al. for the persistence with our model for the ramp effect
   should be possible and would result in a powerful, general, and
   broadly applicable model for charge trapping in WFC3/IR, but it is
   beyond the scope of our study.

\subsection{ The extension of RECTE to other detectors}
Because many IR detectors used in astronomy are manufactured using the
similar technology as \WFC, in principle, RECTE can be extended to
other detectors by adjusting the model parameters. Specifically, the
two near-infrared instruments on board the James Webb Space Telescope
(\textit{JWST}), \textit{NIRCam}, and \textit{NIRSpec} have very
similar focal plane arrays as that of \WFC, and produced by the same
manufacturer (Teledyne Scientific \& Imaging). \textit{NIRCam} has two
identical branches, each of which contains two
$2\mathrm{k}\times2\mathrm{k}$ HgCdTe arrays that cover the wavelength
ranges of 0.6$\mu$m--2.4$\mu$m and 2.4$\mu$m--5.0$\mu$m
\citep{Burriesci2005}. The focal plane array of \textit{NIRSpec}
consists of two butted HgCdTe HAWAII-2RG sensor chip assemblies
\citep{Bagnasco2007}. \edit1{Furthermore, the NIRCam detectors showed
  a persistence trend that is similar to that seen in
  \citet{Smith2008a, Smith2008b}, but with a more complex behavior
  \citep{Leisenring2016}.} The extension of our model to NIRCam may
improve the accuracy of time-resolved observations and save the
valuable \textit{JWST} time by alleviating the need to wait for the
detector state to ``settle.''

With its larger photon collecting area and higher sensitivity compared
to \WFC, \textit{JWST} instruments create new opportunities for
exoplanet atmosphere studies, and are expected to devote large
fractions of available time for time domain exoplanet
observations. For example, with \textit{NIRCam}, it is expected that
with a single transit observation, a signal-to-noise level of 55 at
spectral resolution of $R\sim55$ can be achieved
\citep{Beichman2014}. With such high-quality observations, faint
spectral features may be visible even for GJ1214b, where high-altitude
hazes suppress spectral features. However, the performance of
\textit{JWST} may be limited by an inadequate understanding of the
systematics effect, and it is likely that JWST/NIRCAM may also show a
similar ramp effect. In that case, The extension of RECTE may be used
to solve potential ramp effect systematics for \textit{JWST/NIRCam}
and \textit{JWST/NIRSpec} to obtain high accuracy and efficiency
observations.

\section{Summary}
We present a physically motivated model RECTE to correct the
\WFC ramp effect, the most significant systematic effect in
time-resolved observations with \WFC. We created the model based on
the theory of charge carrier trapping in HgCdTe arrays, and
quantitatively constrained and tested our model by using more than 120
orbits of archival \WFC time-resolved observations. Our model enables
the high fidelity correction of the \WFC ramp effect for observations
taken in both staring and scanning modes regardless of the instrument
setup.  Light curves corrected by RECTE have their residual
photometric systematic errors reduced by a factor of 10, and are
nearly photon noise limited.

Our model can be easily implemented in existing light curve analysis
pipelines. We demonstrate the use of RECTE on two
astrophysically relevant examples and discuss the results of the
observations. The two examples are transiting exoplanet spectroscopy,
and brown dwarf rotational phase mapping, the two most commonly used
time-resolved observational techniques with \WFC.

Existing and future time-resolved observations with \WFC will benefit
from our model and will provide better systematics correction and
improve on-orbit observing efficiency.
\begin{enumerate}
\item Our model does not require a flat baseline to separate the
  instrument-related variability from the variability that is intrinsic
  to the target.
  Therefore, compared to data-driven methods that cannot
    separate instrumental and astrophysical signals, our model provides
    superior ramp effect correction for observations for which flat
    baselines are not available.  Observations of transiting exoplanet
  phase curves and direct photometric/spectroscopic rotational phase
  mapping immediately benefit from our model.
\item For uneven illuminations, e.g., a strong absorbing feature in part
  of the spectral image, different count levels lead to different ramp
  profiles. In the empirical correction, the pixels that are used to
  calculate the correction terms do not have the same count levels as
  the pixels that are corrected. For example, in transit spectroscopy,
  the correction term obtained from white light curves may not
  accurately correct the ramp profile in wavelength channels where the
  illumination level is different from the average level of the white light
  curve. Our model provides a more accurate correction, especially in
  cases where the planet-hosting stars have significant spectral
  features within the wavelength range of the observations.
\item Our model corrects the ramp effect in the first orbit of each
  \HST visit just as well as in the rest of the orbits. In future
  observations, the first orbit light curves no longer have to be
  excluded from the analysis. \glenn{For the most used transit
    observation configuration of 4 continuous orbits per transit, our
    model will increase the \HST observing efficiency by 25\% by alleviating
    the need of discarding the first orbit.}  
\item Our model can be easily extended to similar detectors if
    similar effect is observed. The extensions of our model to
  \textit{NIRCam/JWST} and \textit{NIRSpec/JWST} are expected to
  improve the accuracy and efficiency of future \textit{JWST}
  observations.
\end{enumerate}

\acknowledgments We would like to thank the referee Knox Long for
providing a thorough report, which led to considerable improvement of
the manuscript. We would like to thank Peter McCullough for discussion
on \WFC detectors and comments on the manuscript, Tom Evans, David
Sing, Diana Dragomir, and Hannah Wakeford for helpful discussions on
\WFC transiting exoplanet observations, Laura Kreidberg for
clarifications of GJ1214 data reduction, Marcia Rieke for discussion
on JWST/NIRCam, and Carmen Ortiz Henley for editorial help. Y.Z. is
supported in part by the NASA Earth and Space Science Fellowship
Program - Grant ``NNX16AP54H'', and the Technology Research Initiative
Fund (TRIF) Imaging Fellowship, University of
Arizona. D.A. acknowledges support by the National Aeronautics and
Space Administration under agreement No. NNX15AD94G for the program
Earths in Other Solar Systems. B.W.P.L is supported in part by the
Technology Research Initiative Fund (TRIF) Imaging Fellowship,
University of Arizona. Support for program number 12314, 13418, and
14241 were provided by NASA through a grant from the Space Telescope
Science Institute, which is operated by the Association of
Universities for Research in Astronomy, Inc., under NASA contract
NAS5-26555.

\software{Numpy\&Scipy \citep{VanderWalt2011}, Matplotlib \citep{Hunter2007}, IPython \citep{Perez2007}, Astropy \citep{Robitaille2013}, emcee \citep{Foreman-Mackey2013}, corner.py \citep{Forman-Mackey2016}}
\bibliographystyle{aasjournal}

\begin{thebibliography}{}
\expandafter\ifx\csname natexlab\endcsname\relax\def\natexlab#1{#1}\fi
\providecommand{\url}[1]{\href{#1}{#1}}

\bibitem[{Agol {et~al.}(2010)Agol, Cowan, Knutson, Deming, Steffen, Henry, \&
  Charbonneau}]{Agol2010}
Agol, E., Cowan, N.~B., Knutson, H.~A., {et~al.} 2010, ApJ, 721, 1861.
\newblock \url{http://adsabs.harvard.edu/abs/2010ApJ...721.1861A}

\bibitem[{Anderson {et~al.}(2014)Anderson, Regan, Valenti, \&
  Bergeron}]{Anderson}
Anderson, R.~E., Regan, M., Valenti, J., \& Bergeron, E. 2014

\bibitem[{Apai {et~al.}(2013)Apai, Radigan, Buenzli, Burrows, Reid, \&
  Jayawardhana}]{Apai2013}
Apai, D., Radigan, J., Buenzli, E., {et~al.} 2013, ApJ, 768, 121.
\newblock \url{http://iopscience.iop.org/0004-637X/768/2/121/article/}

\bibitem[{Baggett {et~al.}(2008)Baggett, Hill, Kimble, MacKenty, Waczynski,
  Bushouse, Boehm, Bond, Brown, Collins, {et~al.}}]{Baggett2008}
Baggett, S., Hill, R., Kimble, R., {et~al.} 2008, in SPIE Astronomical
  Telescopes+ Instrumentation, International Society for Optics and Photonics,
  70211Q--70211Q

\bibitem[{Bagnasco {et~al.}(2007)Bagnasco, Kolm, Ferruit, Honnen, Koehler,
  Lemke, Maschmann, Melf, Noyer, Rumler, {et~al.}}]{Bagnasco2007}
Bagnasco, G., Kolm, M., Ferruit, P., {et~al.} 2007, in Proc. SPIE, Vol. 6692,
  66920M

\bibitem[{{Bean} {et~al.}(2010){Bean}, {Miller-Ricci Kempton}, \&
  {Homeier}}]{Bean2010}
{Bean}, J.~L., {Miller-Ricci Kempton}, E., \& {Homeier}, D. 2010, \nat, 468,
  669

\bibitem[{{Bean} {et~al.}(2011){Bean}, {D{\'e}sert}, {Kabath}, {Stalder},
  {Seager}, {Miller-Ricci Kempton}, {Berta}, {Homeier}, {Walsh}, \&
  {Seifahrt}}]{Bean2011}
{Bean}, J.~L., {D{\'e}sert}, J.-M., {Kabath}, P., {et~al.} 2011, \apj, 743, 92

\bibitem[{Beichman {et~al.}(2014)Beichman, Benneke, Knutson, Smith, Lagage,
  Dressing, Latham, Lunine, Birkmann, Ferruit, Giardino, Kempton, Carey, Krick,
  Deroo, Mandell, Ressler, Shporer, Swain, Vasisht, Ricker, Bouwman,
  Crossfield, Greene, Howell, Christiansen, Ciardi, Clampin, Greenhouse,
  Sozzetti, Goudfrooij, Hines, Keyes, Lee, McCullough, Robberto, Stansberry,
  Valenti, Rieke, Rieke, Fortney, Bean, Kreidberg, Ehrenreich, Deming, Albert,
  Doyon, \& Sing}]{Beichman2014}
Beichman, C., Benneke, B., Knutson, H., {et~al.} 2014, Publ. Astron. Soc.
  Pacific, 126, 1134.
\newblock \url{http://iopscience.iop.org/article/10.1086/679566}

\bibitem[{Berta {et~al.}(2012)Berta, Charbonneau, D{\'{e}}sert, {Miller-Ricci
  Kempton}, McCullough, Burke, Fortney, Irwin, Nutzman, \& Homeier}]{Berta2012}
Berta, Z.~K., Charbonneau, D., D{\'{e}}sert, J.-M., {et~al.} 2012, ApJ, 747,
  35.
\newblock \url{http://iopscience.iop.org/0004-637X/747/1/35/article/}

\bibitem[{Buenzli {et~al.}(2014)Buenzli, Apai, Radigan, Reid, \&
  Flateau}]{Buenzli2014}
Buenzli, E., Apai, D., Radigan, J., Reid, I.~N., \& Flateau, D. 2014, ApJ, 782,
  77.
\newblock
  \url{http://stacks.iop.org/0004-637X/782/i=2/a=77?key=crossref.f87581336e83f25f6636ed1f732249df}

\bibitem[{Buenzli {et~al.}(2015)Buenzli, Saumon, Marley, Apai, Radigan, Bedin,
  Reid, \& Morley}]{Buenzli2015}
Buenzli, E., Saumon, D., Marley, M.~S., {et~al.} 2015, ApJ, 798, 127.
\newblock \url{http://iopscience.iop.org/0004-637X/798/2/127/article/}

\bibitem[{Buenzli {et~al.}(2012)Buenzli, Apai, Morley, Flateau, Showman,
  Burrows, Marley, Lewis, \& Reid}]{Buenzli2012}
Buenzli, E., Apai, D., Morley, C.~V., {et~al.} 2012, ApJ, 760, L31.
\newblock \url{http://stacks.iop.org/2041-8205/760/i=2/a=L31}

\bibitem[{{Burriesci}(2005)}]{Burriesci2005}
{Burriesci}, L.~G. 2005, in Society of Photo-Optical Instrumentation Engineers
  (SPIE) Conference Series, Vol. 5904, Cryogenic Optical Systems and
  Instruments XI, ed. J.~B. {Heaney} \& L.~G. {Burriesci}, 21--29

\bibitem[{Charbonneau {et~al.}(2009)Charbonneau, Berta, Irwin, Burke, Nutzman,
  Buchhave, Lovis, Bonfils, Latham, Udry, Murray-Clay, Holman, Falco, Winn,
  Queloz, Pepe, Mayor, Delfosse, \& Forveille}]{Charbonneau2009}
Charbonneau, D., Berta, Z.~K., Irwin, J., {et~al.} 2009, Nature, 462, 891.
\newblock \url{http://www.nature.com/doifinder/10.1038/nature08679}

\bibitem[{Deming {et~al.}(2013)Deming, Wilkins, McCullough, Burrows, Fortney,
  Agol, Dobbs-Dixon, Madhusudhan, Crouzet, Desert, Gilliland, Haynes, Knutson,
  Line, Magic, Mandell, Ranjan, Charbonneau, Clampin, Seager, \&
  Showman}]{Deming2013}
Deming, D., Wilkins, A., McCullough, P., {et~al.} 2013, ApJ, 774, 95.
\newblock \url{http://iopscience.iop.org/0004-637X/774/2/95/article/}

\bibitem[{Dressel(2016)}]{Dressel2016}
Dressel, L. 2016, Wide Field Camera 3, HST Instrument Handbook, 1

\bibitem[{Foreman-Mackey(2016)}]{Forman-Mackey2016}
Foreman-Mackey, D. 2016, The Journal of Open Source Software, 24,
  doi:10.21105/joss.00024.
\newblock \url{http://dx.doi.org/10.5281/zenodo.45906}

\bibitem[{{Foreman-Mackey} {et~al.}(2013){Foreman-Mackey}, {Hogg}, {Lang}, \&
  {Goodman}}]{Foreman-Mackey2013}
{Foreman-Mackey}, D., {Hogg}, D.~W., {Lang}, D., \& {Goodman}, J. 2013, \pasp,
  125, 306

\bibitem[{{Fraine} {et~al.}(2013){Fraine}, {Deming}, {Gillon}, {Jehin},
  {Demory}, {Benneke}, {Seager}, {Lewis}, {Knutson}, \&
  {D{\'e}sert}}]{Fraine2013}
{Fraine}, J.~D., {Deming}, D., {Gillon}, M., {et~al.} 2013, \apj, 765, 127

\bibitem[{Hunter(2007)}]{Hunter2007}
Hunter, J.~D. 2007, Comput. Sci. Eng., 9, 90.
\newblock \url{http://ieeexplore.ieee.org/document/4160265/}

\bibitem[{Kreidberg {et~al.}(2014)Kreidberg, Bean, D{\'{e}}sert, Benneke,
  Deming, Stevenson, Seager, Berta-Thompson, Seifahrt, \&
  Homeier}]{Kreidberg2014}
Kreidberg, L., Bean, J.~L., D{\'{e}}sert, J.-M., {et~al.} 2014, Nature, 505,
  69.
\newblock \url{http://dx.doi.org/10.1038/nature12888}

\bibitem[{Lew {et~al.}(2016)Lew, Apai, Zhou, Schneider, Burgasser, Karalidi,
  Yang, Marley, Cowan, Bedin, Metchev, Radigan, \& Lowrance}]{Lew2016}
Lew, B. W.~P., Apai, D., Zhou, Y., {et~al.} 2016, ApJ,
  doi:10.3847/2041-8205/829/2/L32

\bibitem[{Long {et~al.}(2015{\natexlab{a}})Long, Baggett, \&
  Mackenty}]{Long2015a}
Long, K.~S., Baggett, S.~M., \& Mackenty, J.~W. 2015{\natexlab{a}}, STScI
  Instrument Science Report WFC3, 15

\bibitem[{Long {et~al.}(2015{\natexlab{b}})Long, Baggett, \&
  Mackenty}]{Long2015b}
---. 2015{\natexlab{b}}, STScI Instrument Science Report WFC3, 16

\bibitem[{Long {et~al.}(2014)Long, Baggett, MacKenty, \& McCullough}]{Long2014}
Long, K.~S., Baggett, S.~M., MacKenty, J.~W., \& McCullough, P.~M. 2014, STScI
  Instrument Science Report WFC3, 14, 2014

\bibitem[{Long {et~al.}(2012)Long, Baggett, MacKenty, \& Riess}]{Long2012}
Long, K.~S., Baggett, S.~M., MacKenty, J.~W., \& Riess, A.~G. 2012
  (International Society for Optics and Photonics), 84421W.
\newblock
  \url{http://proceedings.spiedigitallibrary.org/proceeding.aspx?doi=10.1117/12.926778}

\bibitem[{McCullough \& MacKenty(2012)}]{mccullough2012}
McCullough, P., \& MacKenty, J. 2012, STScI Instrument Science Report WFC3, 8,
  2012

\bibitem[{Metchev {et~al.}(2015)Metchev, Heinze, Apai, Flateau, Radigan,
  Burgasser, Marley, Artigau, Plavchan, \& Goldman}]{Metchev2015}
Metchev, S.~A., Heinze, A., Apai, D., {et~al.} 2015, ApJ, 799, 154.
\newblock \url{http://iopscience.iop.org/0004-637X/799/2/154/article/}

\bibitem[{Perez \& Granger(2007)}]{Perez2007}
Perez, F., \& Granger, B.~E. 2007, Comput. Sci. Eng., 9, 21.
\newblock \url{http://ieeexplore.ieee.org/document/4160251/}

\bibitem[{{Reid} {et~al.}(2006){Reid}, {Lewitus}, {Allen}, {Cruz}, \&
  {Burgasser}}]{Reid2006}
{Reid}, I.~N., {Lewitus}, E., {Allen}, P.~R., {Cruz}, K.~L., \& {Burgasser},
  A.~J. 2006, \aj, 132, 891

\bibitem[{{Rieke}(2012)}]{Rieke2012}
{Rieke}, G.~H. 2012, {Measuring the Universe}

\bibitem[{Robitaille {et~al.}(2013)Robitaille, Tollerud, Greenfield,
  Droettboom, Bray, Aldcroft, Davis, Ginsburg, Price-Whelan, Kerzendorf,
  Conley, Crighton, Barbary, Muna, Ferguson, Grollier, Parikh, Nair,
  G{\"{u}}nther, Deil, Woillez, Conseil, Kramer, Turner, Singer, Fox, Weaver,
  Zabalza, Edwards, {Azalee Bostroem}, Burke, Casey, Crawford, Dencheva, Ely,
  Jenness, Labrie, Lim, Pierfederici, Pontzen, Ptak, Refsdal, Servillat, \&
  Streicher}]{Robitaille2013}
Robitaille, T.~P., Tollerud, E.~J., Greenfield, P., {et~al.} 2013, A{\&}A, 558,
  A33.
\newblock \url{http://www.aanda.org/10.1051/0004-6361/201322068}

\bibitem[{Smith {et~al.}(2008{\natexlab{a}})Smith, Zavodny, Rahmer, \&
  Bonati}]{Smith2008a}
Smith, R.~M., Zavodny, M., Rahmer, G., \& Bonati, M. 2008{\natexlab{a}}
  (International Society for Optics and Photonics), 70210J.
\newblock
  \url{http://proceedings.spiedigitallibrary.org/proceeding.aspx?doi=10.1117/12.789372}

\bibitem[{Smith {et~al.}(2008{\natexlab{b}})Smith, Zavodny, Rahmer, \&
  Bonati}]{Smith2008b}
Smith, R.~M., Zavodny, M., Rahmer, G., \& Bonati, M. 2008{\natexlab{b}}
  (International Society for Optics and Photonics), 70210K.
\newblock
  \url{http://proceedings.spiedigitallibrary.org/proceeding.aspx?doi=10.1117/12.789619}

\bibitem[{Swain {et~al.}(2013)Swain, Deroo, Tinetti, Hollis, Tessenyi, Line,
  Kawahara, Fujii, Showman, \& Yurchenko}]{Swain2013}
Swain, M., Deroo, P., Tinetti, G., {et~al.} 2013, Icarus, 225, 432.
\newblock
  \url{http://www.sciencedirect.com/science/article/pii/S0019103513001632}

\bibitem[{van~der Walt {et~al.}(2011)van~der Walt, Colbert, \&
  Varoquaux}]{VanderWalt2011}
van~der Walt, S., Colbert, S.~C., \& Varoquaux, G. 2011, Comput. Sci. Eng., 13,
  22.
\newblock \url{http://ieeexplore.ieee.org/document/5725236/}

\bibitem[{Varley {et~al.}(2015)Varley, Tsiaras, \& Karpouzas}]{Varley2015}
Varley, R., Tsiaras, A., \& Karpouzas, K. 2015, arXiv:1511.09108.
\newblock \url{http://arxiv.org/abs/1511.09108}

\bibitem[{Wakeford {et~al.}(2016)Wakeford, Sing, Evans, Deming, \&
  Mandell}]{Wakeford2016}
Wakeford, H.~R., Sing, D.~K., Evans, T., Deming, D., \& Mandell, A. 2016, The
  Astrophysical Journal, 819, 10.
\newblock
  \url{http://stacks.iop.org/0004-637X/819/i=1/a=10?key=crossref.ff1f558b2e534541a212ec39dc80561a}

\bibitem[{Zhou {et~al.}(2016)Zhou, Apai, Schneider, Marley, \&
  Showman}]{Zhou2016}
Zhou, Y., Apai, D., Schneider, G.~H., Marley, M.~S., \& Showman, A.~P. 2016,
  doi:10.3847/0004-637X/818/2/176.
\newblock \url{https://ui.adsabs.harvard.edu/{\#}abs/2016ApJ...818..176Z}

\end{thebibliography}

\end{document}